\documentclass[fleqn,usenatbib]{mnras}

\usepackage[T1]{fontenc}

\DeclareRobustCommand{\VAN}[3]{#2}
\let\VANthebibliography\thebibliography
\def\thebibliography{\DeclareRobustCommand{\VAN}[3]{##3}\VANthebibliography}

\usepackage{graphicx}	
\usepackage{amsmath}	
\usepackage{newtxtext,newtxmath}

\renewcommand{\vec}[1]{\mbox{\boldmath$#1$}}

\title[Cluster weak-lensing mass bias and scatter]{Calibration of bias and scatter involved in cluster mass measurements using optical weak gravitational lensing}

\author[S. Grandis \& S. Bocquet et al.]
{Sebastian Grandis,$^{1,2}$\thanks{E-mail: s.grandis@physik.lmu.de}
Sebastian Bocquet,$^{1,2}$
Joseph J. Mohr,$^{1,2,3}$
Matthias Klein,$^{1,3}$ 
\newauthor
and Klaus Dolag$^{1,2,4}$
\\
$^{1}$Faculty of Physics, Ludwig-Maximilians-Universit{\"a}t, Scheinerstr. 1, 81679 Munich, Germany\\
$^{2}$Excellence Cluster ORIGINS, Boltzmannstr. 2, 85748 Garching, Germany\\
$^3$Max Planck Institute for Extraterrestrial Physics, Giessenbachstr. 1, 85748 Garching, Germany\\
$^4$Max Planck Institute for Astrophysics, Karl-Schwarzschild-Strasse 1, 85748 Garching, Germany
}

\pubyear{2021}

\hypersetup{draft}
\begin{document}
\label{firstpage}
\pagerange{\pageref{firstpage}--\pageref{lastpage}}
\maketitle

\begin{abstract}
Cosmological inference from cluster number counts is systematically limited by the accuracy of the mass calibration, i.e. the empirical determination of the mapping between cluster selection observables and halo mass. In this work we demonstrate a method to quantitatively determine the
bias and
uncertainties in weak-lensing mass calibration. To this end, we extract a library of projected matter density profiles from hydrodynamical simulations. Accounting for shear bias
and noise,
photometric redshift uncertainties, mis-centering, cluster member contamination, cluster morphological diversity, and line-of-sight projections, we produce a library of shear profiles. Fitting a one-parameter model to these profiles, we extract the so-called \emph{weak  lensing mass} $M_\text{WL}$. Relating the weak-lensing mass to the halo mass from gravity-only simulations with the same initial conditions as the hydrodynamical simulations allows us to estimate the impact of hydrodynamical effects on cluster number counts experiments. Creating new shear libraries for $\sim$1000 different realizations of the systematics, provides a distribution of the parameters of the weak-lensing to halo mass relation,
reflecting their systematic uncertainty. This result can be used as a prior for cosmological inference. We also discuss the impact of the inner fitting radius on the accuracy, and determine the outer fitting radius necessary to exclude the signal from neighboring structures. Our
method is currently being applied to different Stage~III lensing surveys, and can easily be extended to Stage~IV lensing surveys.
\end{abstract}




\section{Introduction}

Measurements of the abundance of massive halos and the galaxy clusters they host are a powerful cosmological probe \citep[e.g.][]{haiman01,allen11, kravtsovborgani12, WEINBERG2013}. However, the derived constraints are systematically limited by the uncertainty in the relationship between the underlying halo mass and the (multi-)observable halo properties \citep[for a review of cluster mass measurement techniques, see][]{pratt19}. Of these observables, weak gravitational lensing is particularly promising because it traces the total enclosed mass and is therefore insensitive to the complex halo dynamics and to the state of the intra-cluster medium. Indeed, weak lensing mass calibration has been established as the method of choice for a robust determination of the cluster mass scale \citep{hoekstra12, applegateetal14, vonderlinden14, schrabback18a, murata19, dietrich19, mcclintock19}. Various cluster samples have been used for cosmological studies with a joint weak-lensing mass calibration \citep{Mantz15, bocquet19, des_y1_cluster}.

The weak-lensing shear measurement of a galaxy cluster consists of the tangential distortion of the shape of background source galaxies and an estimate of the source redshift distribution \citep[for a didactic review of cluster weak lensing, see][]{umetsu20}. Several sources of  systematic and statistical uncertainty impact the mapping between the tangential ellipticity profile and halo mass. The most prominent sources of statistical noise are:
\begin{itemize}
    \item the morphological heterogeneity of the cluster mass distribution (such as the different concentration, orientation, halo shape and substructure),
    \item the varying amount of line-of-sight projection by nearby
    correlated structures,
    \item the contribution of uncorrelated large-scale structure along the line of sight,
    \item the noise in the shape measurements of source galaxies due to the dispersion of the intrinsic ellipticities, and
    \item the lack of knowledge of the actual cluster center and its displacement w.r.t. the observationally chosen center.
\end{itemize}
The major sources of systematic uncertainties are
\begin{itemize}
    \item the accuracy of the hydrodynamical simulation predictions of the matter distribution of halos,
    \item the accuracy of the shape measurements (especially in crowded environments and in the limit of large amounts of shear),
    \item the accuracy of the photometric redshift estimates used for the source galaxy redshift distribution,
    \item the ability to constrain the distribution of offsets between the observationally chosen centers and the true cluster center (i.e. the mis-centering distribution), and 
    \item the amount of cluster member contamination of the true lensing signal.
\end{itemize}

Typically, the modelling of the weak-lensing shear profile as a function of halo mass is anchored around the fact that the average halo mass profile is well described by a Navarro-Frenk-White density profile \citep[NFW,][]{NFW}. Conveniently, the gravitational shear caused by an NFW halo can be calculated analytically \citep{bartelmann96}. However, as discussed, real halos do not exhibit perfect, spherically symmetric NFW mass profiles and they are not perfectly isolated objects. Therefore, the inferred mass from this approach -- we refer to this mass as the weak-lensing mass $M_\mathrm{WL}$ -- is a noisy and biased estimator of the actual halo mass $M_\mathrm{halo}$. Numerical simulations are then used to calibrate this noise and bias as a function of halo mass and redshift \citep[e.g.][]{becker11, Oguri2011MNRAS.414.1851O, Bahe12, lee18}. For completeness, we also note that a parallel approach has recently emerged where the halo mass profile, parametrized as the halo--matter correlation function, is predicted using emulation techniques \citep{Nishimichi2019ApJ...884...29N}.

For practical analyses of weak-lensing data, the generic calibrations of the $M_\mathrm{WL}$--$M_\mathrm{halo}$ relation (hereafter weak-lensing to halo mass relation) discussed above are of limited use. Most importantly, these calibrations do not account for the fact that only observationally determined cluster centers are available, and these scatter around the true halo center \citep[e.g.][]{lin04a,saro15, zhang17}. However, the above mentioned simulation-based calibrations assume perfect centering, and the established relations cannot be adapted to account for mis-centered shear profiles, because calculating the shear signal from a mis-centered matter density profile is not equivalent to computing the shear profile of a centered matter density profile and then estimating the effect of mis-centering by azimuthal averaging. Instead, the effect of mis-centering must be applied at the level of the simulations. Then, all other sources of uncertainty listed above could in principle be explicitly marginalized over in the cosmological analysis. We attempt this approach and update the cosmological pipeline used for the latest analysis of the South Pole Telescope-selected (SPT) cluster sample with targeted weak-lensing measurements \citep{bocquet19} to explicitly marginalise over all these effects in a full forward modelling approach. We find that this procedure is computationally expensive, even for cluster samples with a few hundred weak-lensing measurements.

We therefore propose to fit a simple model for the shear profile (a projected NFW profile with an effective contribution from mis-centering) to the data while explicitly accounting for shape noise. All other sources of statistical and systematic uncertainties are folded into the weak-lensing to halo mass relation. This treatment allows us to capture all relevant effects while reducing the computational burden on the cosmological analysis pipeline (e.g. for the analysis of SPT clusters with weak-lensing mass calibration using data from the Dark Energy Survey, Bocquet et al. in prep.). We note that a similar approach has been performed in \cite{Mantz15, bocquet19, dietrich19, Schrabback2020arXiv200907591S}.

A systematic uncertainty in cluster lensing that is receiving increasing attention is caused by the impact of baryonic feedback effects on the halo matter profile \citep[e.g.][]{Debackere2021arXiv210107800D}. In the presence of these effects, only hydrodynamical simulations with carefully tuned sub-resolution physics models can be expected to produce realistic halo profiles. However, to perform a cluster cosmology study, the halo profile needs to be related to a halo mass which is linked to cosmology via the halo mass function. In principle, this could be addressed by establishing both the calibration of the weak-lensing to halo mass relation \citep[as in, e.g. ][]{lee18} and of the halo mass function on hydrodynamical simulations \citep[as in, e.g.][]{bocquet16}. 
However, one reason not do so is that the advent of emulators based on suites of gravity-only simulations has dramatically increased the accuracy with which the mass function can be predicted for a given cosmological model \citep{mcclintock19b, Nishimichi2019ApJ...884...29N, Bocquet2020ApJ...901....5B}. To be able to benefit from these emulators, we propose a novel approach: we keep the halo mass function as defined by gravity-only halo masses. Then, we link these gravity-only masses to realistic halo profiles as obtained from hydrodynamical simulations. This is achieved by using pairs of hydrodynamical and gravity-only simulations with identical initial conditions. As a result, in our approach, the impact of hydrodynamical effects is folded into the weak-lensing to halo mass relation along with the other systematic and statistical effects already discussed. Repeating this exercise with different hydrodynamical simulations -- ideally weighted by their agreement with real observations -- allows us to quantify the error budget related to uncertainties in the hydrodynamical modeling.

We organize this paper as follows. In Section~\ref{sec:shear_lib} we discuss how we produce realistic tangential shear profiles for halo catalogs by combining data extracted from simulations and calibration products derived from observations. We seek to emulate a typical Stage III weak-lensing survey, taking mainly inspiration from the Dark Energy Survey year 1 data \citep[DES Y1]{desy1_data} and  the cluster catalog selected by the SPT Sunyaev-Zel'dovich effect survey \citep[SPT-SZ;][]{benson13, bleem15, bocquet19}. We then define our model for the shear profile in Section~\ref{sec:shear_model}. We then fit this model to the realistic shear profiles in the library to extract weak-lensing masses.
In Section~\ref{sec:shearmodel&mWL} we then describe how we characterize the relation between weak-lensing and halo mass, introducing the weak-lensing bias and scatter. By varying the input values of the shear library creation, we propagate the uncertainty on the weak-lensing systematics onto the weak-lensing bias and scatter. We demonstrate how this procedure provides a posterior on the weak-lensing bias and scatter in Section~\ref{sec:calib_gen_survey}, which provides a crucial prior for weak-lensing calibrated number count experiments. We then discuss the impact of the inner (Section~\ref{sec:inner_fitting_radius}) and outer (Section~\ref{sec:outer_radius}) fitting radius, as well as other systematic effects, which are harder to quantify (Section~\ref{sec:further_sys}), and compare different approaches to modelling the systematic uncertainty stemming from the modelling of hydrodynamical effects (Section~\ref{sec:hydro_approaches}). Finally, we summarize our findings in the conclusions (Section~\ref{sec:conc}).

\begin{figure*}
	\includegraphics[width=\textwidth]{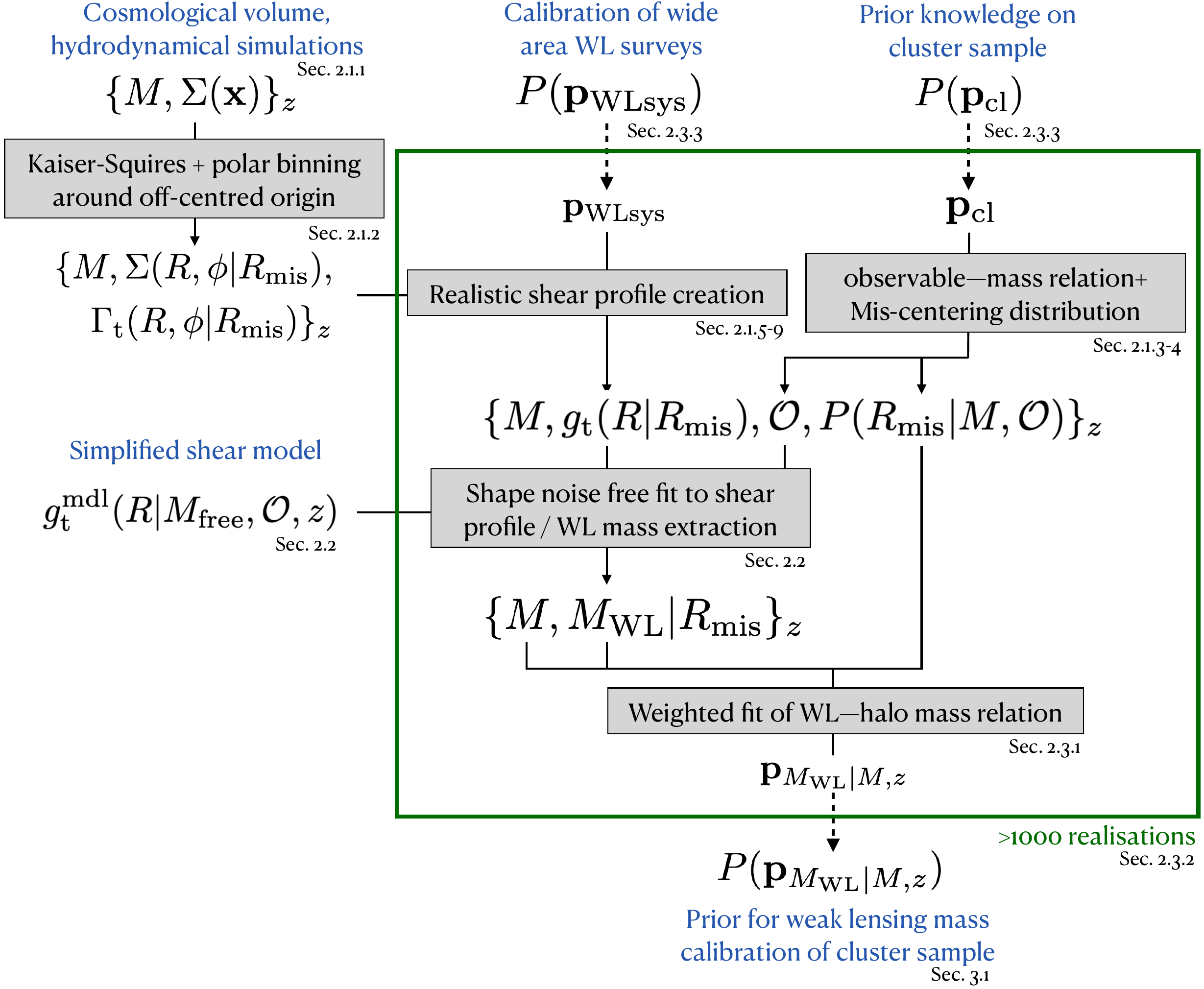}
	\vskip-0.10in
    \caption{Graphic representation of the method used to quantify the accuracy of the weak lensing (WL) mass extraction by combining the surface matter densities $\Sigma(\vec{x})$ around halos of mass $M$ from cosmological volume, hydrodynamical simulations with calibrations of the WL data in wide-area surveys and priors on the cluster sample. By assuming a simplified shear model, these inputs can be compressed into a single posterior on the parameters of the mapping between WL and halo mass. For ease of navigation, we add the sections in which each component of the diagram is discussed.}\label{fig:diagram}
\end{figure*}

\section{Method}

In this section, we first discuss how we create a realistic library of shear profiles for a halo catalog by combining information from hydrodynamical simulations and calibrations of observed data (cf. Section~\ref{sec:shear_lib}). We then discuss the shear model we use to measure associated weak-lensing masses for each of the realistic shear profiles (cf. Section~\ref{sec:shear_model}). Subsequently, we establish the statistical relation between weak-lensing and halo mass (cf. Section~\ref{sec:shearmodel&mWL}). Finally, we present a method to propagate the uncertainties on the calibration products to the parameters of the weak-lensing to halo mass relation, producing a posterior distribution
of these parameters that can be used as a prior in cosmological analyses (cf. Section~\ref{sce:accuracy_WLhalo_rel}).
Fig.~\ref{fig:diagram} is a flow diagram of our method, supplemented with the reference to the sub-section corresponding to each module.

\subsection{Library of realistic shear profiles}\label{sec:shear_lib}

\subsubsection{Simulations}\label{sec:sims}

Our work is mainly based on Box1 of the Magneticum Pathfinder suite of cosmological hydrodynamical simulations (Dolag et al., in prep.).\footnote{\url{http://www.magneticum.org/index.html}} Additional technical aspects of the simulations are available in the literature \citep{Hirschmann2014MNRAS.442.2304H, Teklu2015ApJ...812...29T, Beck2016MNRAS.455.2110B, Dolag2017Galax...5...35D}.
The box size (896~$h^{-1}$ Mpc on a side), the resolution ($2\times1526^3$ particles) and the mass sampling (1.3$\times$10$^{10}\,h^{-1}$ $M_\odot$ for dark matter particles, 2.6$\times$10$^9\,h^{-1}$ $M_\odot$ for gas particles) allow for a good halo resolution down to about $10^{14}\,h^{-1}\,$M$_\odot$ and for sufficiently large numbers of halos. The cosmological parameters match the WMAP7 constraints for a spatially flat $\Lambda$CDM model \citep{Komatsu2011ApJS..192...18K}: $\Omega_m=0.272$, $\Omega_b=0.0457$, $H_0=70.4$, $n_s=0.963$, $\sigma_8=0.809$. In addition to the full-physics hydrodynamical run, a gravity-only run using the same initial conditions was also produced.

\begin{figure}
 \includegraphics[width=\columnwidth]{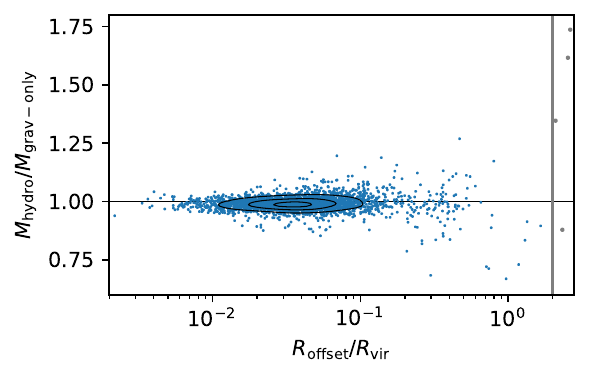}
 \vskip-0.15in
 \caption{\label{fig:mass_ratio}
 Mass ratio hydro/gravity-only as a function of center offset for $z=0.25$. Points show all halos in the snapshot above the mass cut $M_{200c}(\text{gravity-only})>1.56\times10^{14}\,h^{-1}$ M$_\odot$.
 Contours enclose 30\%, 60\%, and 90\% of all points.
 The vertical gray line shows the cut $R_\mathrm{offset}=2 R_\mathrm{vir}$.}
\end{figure}

We use the snapshots 52, 72, 96, 116, 144 which correspond to redshifts 1.18, 0.78, 0.47, 0.25, 0.00. Halos are identified using a modified version of the \textsc{subfind} algorithm using a linking length $b=0.16$ \citep{Springel2001MNRAS.328..726S, Dolag2009MNRAS.399..497D}. Spherical over-density masses are then computed out from the potential minimum of each halo. For each snapshot, we extract halo catalogs from the gravity-only simulation down to $M_{200c}>1.56\times10^{14}\,h^{-1}$ M$_\odot$. With this cut, the region within $r_{200c}$ is comfortably resolved with at least $10^4$ particles of each species on average. For the hydrodynamical run, we set a limit that is lower by a factor of 0.7. Starting from the gravity-only halo catalogs, we assign each halo a counterpart from the hydrodynamical catalogs matching by halo position. To avoid spurious matches, we require that the center offsets are within two times the virial radius of the halo in the gravity-only catalog. This cut removes less than 0.5\% of all halos. The scatter in $M_{200c}$ between both catalogs is within ~30\%; this motivated the choice to extract halos from the hydrodynamical run down to 0.7 of the mass cut in the gravity-only simulation. As an example, we illustrate the matched catalogs for $z=0.25$ in Fig.~\ref{fig:mass_ratio}. In this snapshot, the hydro masses are on average $0.99\pm0.03$ lower than their gravity-only counterparts.

To investigate how much our findings depend on the specifics of the Magneticum simulations, we use the Illustris TNG300-1 simulations as a cross-check \citep{Pillepich2018MNRAS.475..648P, Marinacci2018MNRAS.480.5113M, Springel2018MNRAS.475..676S, Nelson2018MNRAS.475..624N, Naiman2018MNRAS.477.1206N, Nelson2019ComAC...6....2N}. These include $2\times2500^3$ resolution elements for a box size of $302.6$~Mpc on a side. While TNG300-1 features a higher mass resolution than Magneticum Box1 its simulation volume is significantly smaller and therefore only tracks a small number of high-mass halos. The cosmology corresponds to the Planck2015 constraints for a spatially flat $\Lambda$CDM cosmology \citep{PlanckCollaboration2016A&A...594A..13P}: $\Omega_m=0.3089$, $\Omega_b=0.0486$, $\sigma_8=0.8159$, $n_s=0.9667$, and $h=0.6774$. The TNG simulations feature a full-physics hydrodynamical run (TNG300-1) and a gravity-only run (TNG300-1-Dark) that uses the same initial conditions.
Because we are only interested in spot-checking our results obtained from the Magneticum simulations, we restrict ourselves to using the TNG300-1 snapshots 51 and 98, which correspond to redshifts 0.95 and 0.01. Halos are identified via a FOF algorithm and spherical overdensity masses are built out from the halo potential minimum. We extract halos and match the hydrodynamical and gravity-only halo catalogs following the procedures detailed above with the exception that the mass limit for TNG300-1-Dark is set to $M_{200c}>5\times10^{13}\,h^{-1}$ M$_\odot$; the limit for the hydrodynamical TNG300-1 is 0.7 times that value. The cross-check between results obtained from the Magneticum simulations and TNG are presented in Section~\ref{sec:further_sys} and Appendix~\ref{sec:app_A}.

\begin{figure*}
	\includegraphics[width=0.49\textwidth]{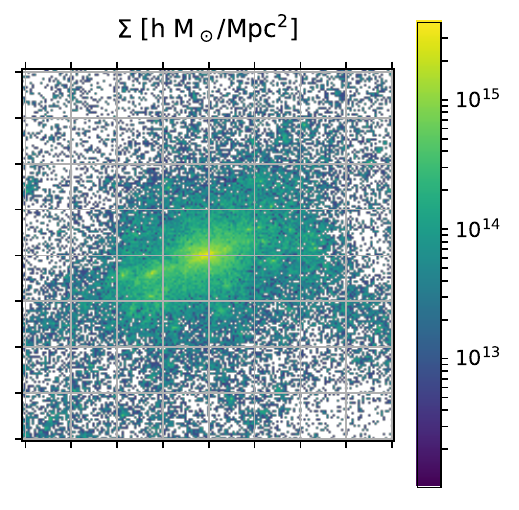}
	\includegraphics[width=0.49\textwidth]{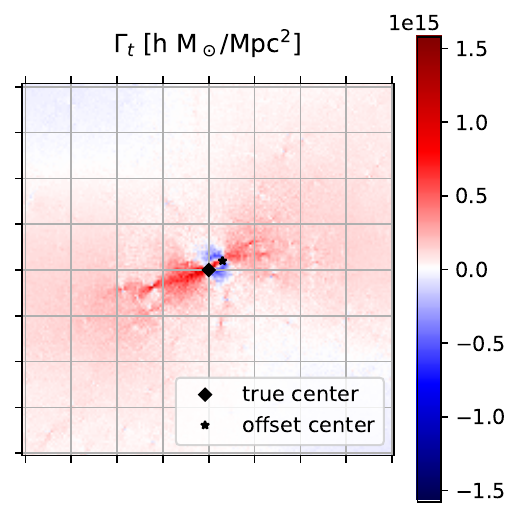}
	\vskip-0.10in
    \caption{Example surface matter density map (left panel) around a halo in the hydrodynamical simulation. The grid denotes 1\,$h^{-1}$Mpc~$\times$\,1\,$h^{-1}$Mpc squares. To capture the impact of correlated structure along the line of sight, we project along $\pm 20$ $h^{-1}$Mpc in the perpendicular direction. After applying the Kaiser-Squires algorithm to obtain scaled versions of the shear components $\Gamma_{1,2}$ along the Cartesian coordinates, we choose isotropically oriented, offset centers (star in right panel), around which we compute the scaled tangential shear $\Gamma_\text{t}$ in the presence of mis-centering.}\label{fig:eg_maps}
\end{figure*}

\begin{figure}
\includegraphics[width=\columnwidth]{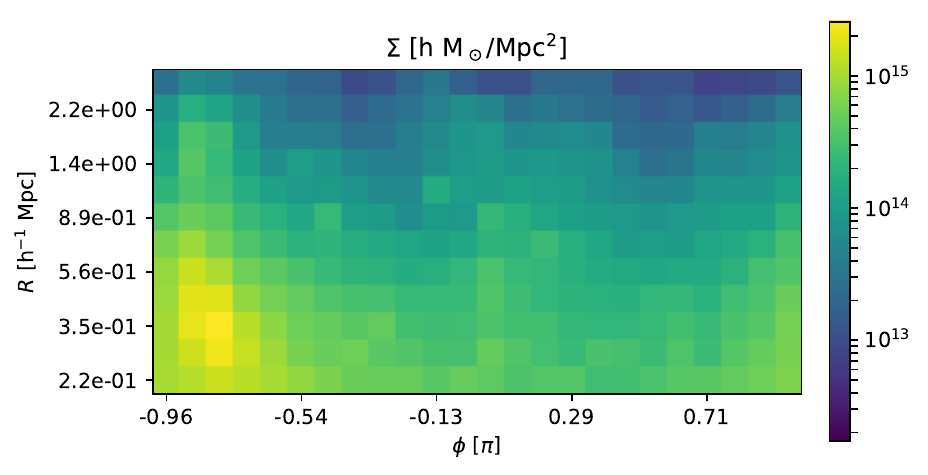}
\includegraphics[width=\columnwidth]{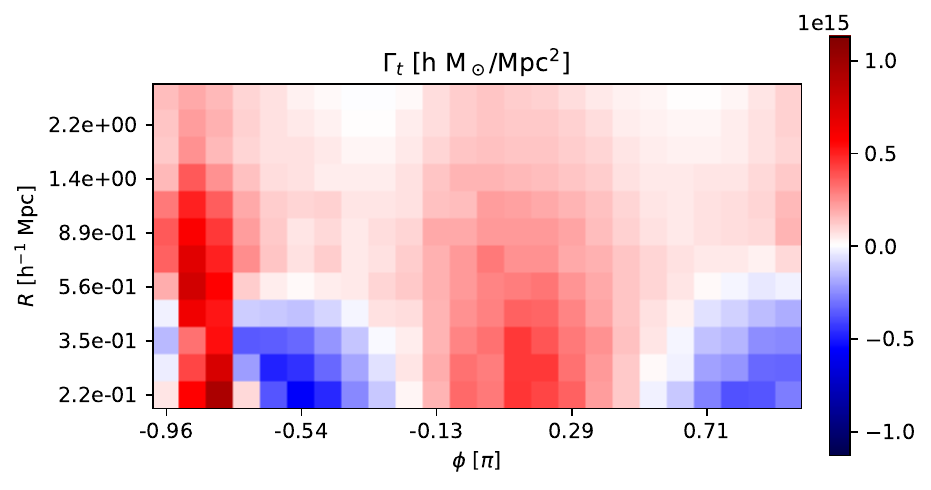}
\caption{To compress the surface density map (upper panel) and the scaled tangetial shear map (lower panel), while conserving information on the azimuthal anisotropy of the maps, we define bins in polar coordinates, equally spaced in log-radius $R$ and azimuthal angle $\phi$ around the offset center. Combined with the critical surface density $\Sigma_\text{crit}$ (cf. Section~\ref{sec:photo-z_bias}), we can compute the reduced shear profile while accounting for the azimuthal anisotropy of surface matter density and the tangential shear.}\label{fig:polar_binning}
\end{figure}


\subsubsection{Convergence and shear maps}\label{sec:matter_profiles}

We now extract cylindrically projected (total) matter density maps centered on the halo potential minimum for the halos in the simulations. 
Due to the steepness of the halo mass function, 
the halo catalogs 
have far more low mass halos.  In our analysis we seek to explore halo mass trends with approximately equal fidelity over a broad range of masses.  Therefore, 
at low halo masses, we randomly sub-sample our halo catalogs such that each mass bin of width $\Delta \log_{10}M_\text{200c}=0.1$ has at most about 100 halos. Because extracting the projected matter maps from the simulations is expensive, this sub-sampling significantly reduces the cost of our analysis, while still ensuring enough statistical constraining power (cf. Section~\ref{sce:accuracy_WLhalo_rel}).

The cylindrical projection depth is set to $l_\text{proj}= \pm 20$ $h^{-1}$Mpc, which is sufficient to capture the effects of 
correlated large-scale structure \citep{becker11}. We bin the map in 20kpc$h^{-1}~\times~$20kpc$h^{-1}$ quadratic bins, providing a good balance between sufficient sampling for the 0.2 $h^{-1}$Mpc minimal scale we are interested in (cf. Section~\ref{sec:inner_fitting_radius}), and memory usage. For each halo, we produce three (almost) independent 
matter density maps by projecting along the three Cartesian coordinate axes. The left panel of Fig.~\ref{fig:eg_maps} shows one of the matter density maps of an exemplary halo. 

For each projection, we then compute the scaled shear components by applying the Kaiser-Squires algorithm \citep{kaisersquires},
\begin{equation}
 \Gamma_i =\mathcal{F}^{-1}\big[ \chi_i(\vec{k}) ~ \mathcal{F}[\Sigma] \big], \text{ with } \vec{\chi}(\vec{k}) = \frac{-1}{|
\vec{k}|^2} \begin{pmatrix} k_1^2 - k_2^2 \\ k_1 k_2 \end{pmatrix}, 
\end{equation}
where $\mathcal{F}[\cdot]$ stands for the 2d Fourier transform, $\mathcal{F}^{-1}[\cdot]$ for the inverse 2d Fourier transform, and $\vec{k}$ is the associated wave vector. To obtain the actual shear map, the scaled shear needs to be divided by the critical density $\Sigma_\text{crit}$ (cf. Section~\ref{sec:photo-z_bias}). Given that we wish to perturb the critical density to account for photometric redshift uncertainties, we save the scaled shear at this stage of the analysis.

We then define 15 offset centers for each projection direction, for each halo, at 15 different mis-centering radii, equally log-spaced between (0.01, 1) $h^{-1}$Mpc. These centers are positioned isotropically around the true center. For each offset center, as well as for the true center, we compute the scaled tangential shear,
\begin{equation}
\Gamma_\text{t} (\vec{x})  = - \Gamma_1 (\vec{x}) \cos(2\phi) - \Gamma_2(\vec{x}) \cos(2 \phi),
\end{equation}
where $\phi$ is the polar angle of $\vec{x}$. An example of such a mis-centered scaled tangential map is shown in the right panel of Fig.~\ref{fig:eg_maps}. Note the pronounced azimuthal anisotropy around the offset center (black star). In directions perpendicular to the mis-centering direction, the tangential shear is negative. This anisotropy needs to be traced to compute the correct reduced shear profile (cf. Section~\ref{sec:photo-z_bias}).

To efficiently handle the large amount of data -- several 100 halos, times 3 projections, times 15+1 centers -- we need to compress the map information in a way that conserves the azimuthal anisotropy. To this end we re-bin the surface mass density maps and the scaled tangential shear maps in radial bins, equally log-spaced between (0.01, 10) $h^{-1}$Mpc, and with 24 bins in azimuthal angle $\phi$. The resulting re-binned maps are shown in Fig.~\ref{fig:polar_binning}. These maps are significantly compressed in comparison to the Cartesian maps, while capturing the relevant information (radial trend, azimuthal anisotropy) needed to produce realistic reduced shear profiles.

\begin{figure*}
	\includegraphics[width=\textwidth]{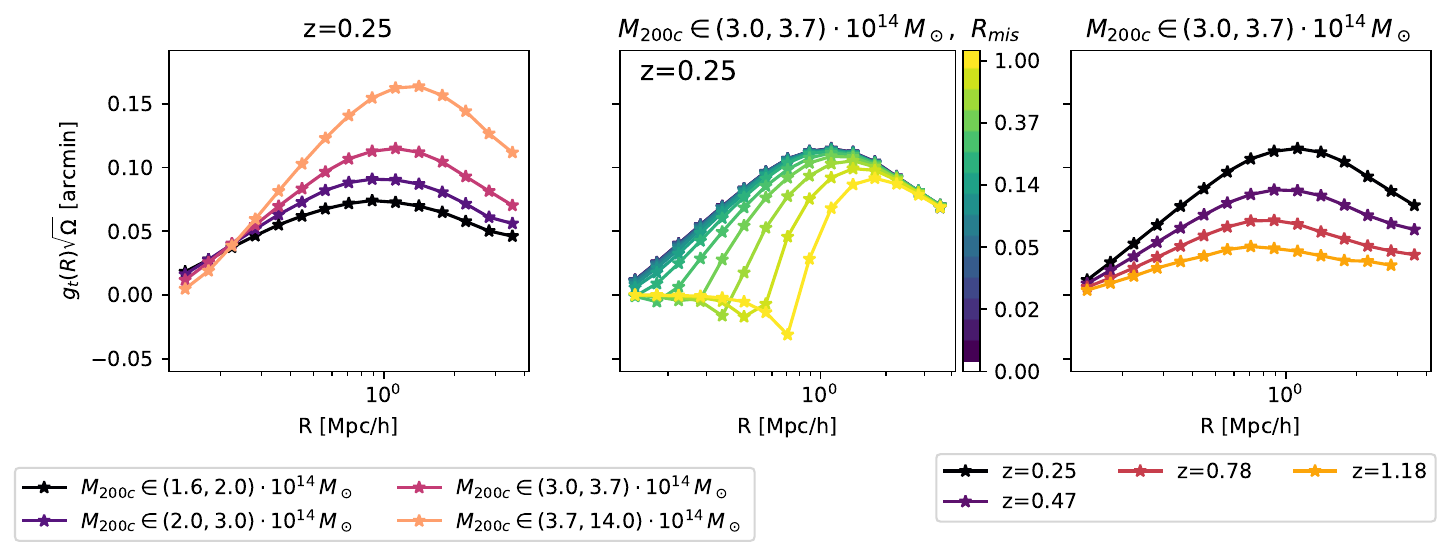}
	\vskip-0.10in
    \caption{Shear profiles stacked in mass bins multiplied with the square root of the area in degrees, visualised for  different dimensions of our shear profile library: mass bins (left panel),  mis-centring radius used for extraction (center), redshift of the snapshots (right). The amplitude of the signal can be compared to the typical shape-noise of $\sim 0.3 / \sqrt{n_\epsilon}$, where $n_\epsilon$ is the source density per arcmin$^2$.
    We construct this shear profile library for a given set of weak-lensing survey specifications and systematics parameters. 
    Varying these parameters allows us to sample the space of weak-lensing systematics.
    }\label{fig:shear_library}
\end{figure*}

To transform the matter profile library into a shear profile library, we must specify several characteristics
of the weak-lensing survey.  These include: photometric redshift bias, shear measurement bias, impact of uncorrelated large-scale structure, cluster member contamination together with the halo catalog cluster observables and mis-centering distribution. The values of these properties are subject to systematic uncertainties, which in turn contribute to the final systematic halo mass uncertainty. We parametrize these additional elements, allowing us to marginalise over these uncertainties. Note also, that the weak-lensing signal is not well defined for $z=0$. We therefore omit that snapshot when considering results based on shear profiles.

We choose the numerical values for these parameters by drawing from the Stage III DES Y1 weak-lensing survey \citep{mcclintock19} and the South Pole Telescope cluster survey \citep{bleem15}. The values 
need to be adjusted to the specific weak-lensing survey and cluster catalog before application in cosmological analyses. 

\subsubsection{Cluster observables}\label{sec:obs_mass}

Many weak-lensing systematics (such as the cluster member contamination and the mis-centering distributions) are empirically calibrated as functions of cluster observables.  We thus need to assign such observables if we want to realistically model the weak-lensing systematics. Following \citet{saro15, bleem19, grandisip}, we assign a richness $\lambda$ by drawing
\begin{equation}
    \ln \lambda \sim \mathcal{N} \big(\langle \ln \lambda \rangle (M_\text{200c}, z); \sigma_\text{tot}^2\big),
\end{equation}
with
\begin{equation}
    \langle \ln \lambda \rangle (M_\text{200c}, z) = \ln A_\lambda + B_\lambda \ln \Big(\frac{M_\text{200c}}{3e14  ~h^{-1}\text{ M}_\odot } \Big) + C_\lambda \ln \Big(\frac{E(z_\text{cl})}{E(0.6)}\Big), 
\end{equation}
where $E(z)$ is the critical density of the Universe in units of the present day critical density, and with
\begin{equation}
    \sigma_\text{tot}^2 = \exp\big(2\ln\sigma_{\lambda}\big) + \frac{\exp\langle \ln \lambda \rangle (M_\text{200c}, z)-1}{\exp (2\langle \ln \lambda \rangle (M_\text{200c}, z))}.
\end{equation}
The parameters $(A_\lambda,\, B_\lambda,\, C_\lambda, \ln\sigma_\lambda)$ parameterize the normalization, mass trend, redshift trend and logarithmic scatter in the $\lambda$-mass relation, whose systematic uncertainties are reflected as parameter uncertainties.

\subsubsection{Mis-centering distribution}\label{sec:mis_centering}

When extracting shear profiles in real data, the chosen center does not coincide with the halo center. 
To properly assess how probable each of our mis-centering radii is, we 
adopt the following mis-centering distribution: 
\begin{equation}\label{eq:P_Rmis}
    P(R_\text{mis}|\lambda) = \rho \text{Rayl}\Big( \frac{R_\text{mis}}{R_\lambda}; \sigma_0\Big) + (1-\rho) \text{Rayl}\Big(\frac{R_\text{mis}}{R_\lambda}; \sigma_1\Big),
\end{equation}
with $R_\lambda = (\lambda/100)^{0.2}\,h^{-1}$ Mpc. This is a two-component Rayleigh distribution that provides a good description of the mis-centering of optical centers such as the central brightest cluster galaxy \citep{saro15, bleem19} with respect to true halo centers. A large fraction $\rho$ of clusters contain well centered objects (typically $\sigma_0<0.1$), and a smaller sub-population of disturbed and thus strongly mis-centered clusters (typically $\sigma_1>0.1$) show large mis-centering effects \citep[see also optical--X-ray studies such as][]{lin04a}. In general, the strength of the mis-centering depends on some cluster observables, such as richness. The parameters of the mis-centering distribution in this case are $(\rho,\, \sigma_{0},\, \sigma_{1})$.

\subsubsection{Photo-z uncertainty in source redshifts}\label{sec:photo-z_bias}
We assume that our generic survey has a source redshift distribution $P(z_\text{s}) = 0.5\,z_\text{s,0}^{-3} z_\text{s}^2 \exp (-z_\text{s}/z_\text{s,0})$, with $z_\text{s,0}=0.2$ \citep[following the parametrisation suggested by][]{smail94}. For a cluster at redshift $z_\text{cl}$ we model the background selection by imposing the cut $z_\text{s}>z_\text{cl}+0.1$. In a realistic case, the real source redshift
distribution of a specific survey would have to be used. Also the background selection method could be different.

Given a source redshift distribution and a background selection we compute the lensing efficiency
\begin{eqnarray}
\label{eq:beta}
    \Sigma_\text{crit}^{-1}(z_\text{cl}) & = & \frac{4\pi G d_\text{A}(z_\text{cl})}{c^2} \Big\langle\frac{d_\text{A}(z_\text{cl}, z_\text{s})}{d_\text{A}(z_\text{s})} \Big\rangle_{z_\text{s}>z_\text{cl}+0.1} \\ \nonumber
    & & \big(1 + \delta_\beta(z_\text{cl})+\alpha_\beta \sigma_\beta(z_\text{cl}) \big),
\end{eqnarray}
where $\delta_\beta(z_\text{cl})\pm \sigma_\beta(z_\text{cl})$ is an estimate of the bias on the lensing efficiency due to photometric redshift measurements, together with the uncertainty on this bias. For a deep photometric survey aiming at measuring cosmic shear, this quantity is one of the most relevant systematics, and thus a natural calibration product \citep[e.g. ][]{hoyle18}. For our generic survey we assume
\begin{equation}
    \delta_\beta(z_\text{cl})= 
\begin{cases}
\begin{split}
    &0.025  &\text{ for }z_\text{cl}<0.7 &\\
    &0.05 - 0.025 \Big(\frac{1+z_\text{cl}}{1.7} \Big)^7 &\text{ for }z_\text{cl}>0.7 & ,
\end{split}
\end{cases}
\end{equation}
and
\begin{equation}
    \sigma_\beta(z_\text{cl}) = 0.02\, \Big(\frac{1+z_\text{cl}^2}{1.49}\Big)^2.
\end{equation}
This prescription qualitatively follows the photometric redshift bias and uncertainty in the DES Y1 data which at $z_\text{cl}\gtrsim 0.7$ grow quickly \citep{mcclintock19}.
The parameter $\alpha_\beta$ is introduced to vary the strength of the photo-z bias within its errors. By construction, its mean value is 0, while its variance is 1.


The lensing efficiency allows us to compute the convergence map, and the tangential shear map,
\begin{equation}\label{eq:std_WL_1}
    \kappa(R, \phi | R_\text{mis}) =  \frac{\Sigma (R, \phi | R_\text{mis})}{\Sigma_\text{crit}(z_\text{cl}) }\text{, and }\gamma_\text{t}(R, \phi | R_\text{mis}) =  \frac{\Gamma_\text{t}(R, \phi | R_\text{mis})}{\Sigma_\text{crit}(z_\text{cl}) },
\end{equation}
while accounting for the azimuthal anisotropy of both maps. Besides the natural anisotropy introduced by the halo ellipticity and morphology even for well centered cases, assuming offset centers strongly contributes to the anisotropy, as can be seen in Fig.~\ref{fig:eg_maps} and Fig.~\ref{fig:polar_binning}.

The reduced shear profile is then obtained by first computing the reduced shear map, accounting for azimuthal anisotropy, and then averaging over azimuthal angles, 
\begin{equation}
  g_\text{t}(R | R_\text{mis}) = \int \text{d}\phi \frac{\gamma_\text{t}(R, \phi | R_\text{mis})  }{1-\kappa(R, \phi | R_\text{mis}) }.
\end{equation}
Inverting this precise order -- reduced shear map first, azimuthal average second -- leads to errors in the reduced shear profile of a few percent (especially around the position of the true center) and to $\le$1\% level shifts 
in the inferred WL bias.

\subsubsection{Uncorrelated large-scale structure}
The next step in the creation of a realistic shear measurement is to add uncorrelated large-scale structure projections to the shear profile. Using the prescription presented by \citet{hoekstra03}, we compute the covariance of the large-scale structure $C_\text{LSS}(R, R^\prime)$.  We model this effect by drawing a Gaussian deviate around the reduced shear profile $\tilde{g}(R | R_\text{mis}) \sim \mathcal{N}(g_\text{t}; C_\text{LSS})$. Importantly, this approach accounts for the line-of-sight projections due to structures that are not correlated with the target halo. This effect depends on the cosmological model through the matter power spectrum and the distance sensitivity of the lensing kernel. It is however independent on the cluster mass. Furthermore, in current Stage III lensing surveys, this effect in much smaller than the shape noise. The correlated structure effects are already captured within the projected matter density profiles extracted from the simulations (cf. Section~\ref{sec:matter_profiles}).

\subsubsection{Multiplicative shear bias}\label{sec:mult_shear_bias}

The measured reduced shear results from the predicted reduced shear after accounting for the multiplicative shear bias introduced by the shape measurements. Following \citet{sheldon17}, we model this as
\begin{eqnarray}
    g_\text{t, meas}(R | R_\text{mis}) &=& \bigg(1+ m + \alpha_\gamma \sigma_m + \alpha_\text{NL}g^2_\text{t}(R | R_\text{mis}) \bigg)  \\ \nonumber
    & & g_\text{t}(R | R_\text{mis}), 
\end{eqnarray}

where $m$ is the multiplicative shear bias, $\sigma_m$ its error. We set $m=\sigma_m=0.01$, which are typical values for Stage III lensing surveys. We introduce the parameter $\alpha_\gamma$ to modulate the strength of the multiplicative shear bias. We also introduce a parameter that accounts for the non-linear shear response $\alpha_\text{NL}$. To date, image simulations used to calibrate the shape measurements focus most attention on accurate calibration of the multiplicative shear bias $m$ and its uncertainty $\sigma_\text{m}$
while little attention is given to the performance of shape measurements in the large shear regime.

\subsubsection{Cluster member contamination}\label{sec:member_cont}

The contamination of the background sample by cluster member galaxies leads to a suppression of the measured shear signal, because the cluster members are not sheared by the cluster potential. Several techniques have been proposed to calibrate this effect empirically, and they all result in a cluster member contamination profile $f_\text{cl}$, which naturally has uncertainties on its parameters \citep[see][and references therein]{varga19}. We choose here the parameterisation 
\begin{equation}
    f_\text{cl}(R|\lambda) = A_\text{fcl} \Big(\frac{\lambda}{67}\Big)^{B_\text{fcl}} \frac{\tilde{f}(x)}{\tilde{f}(1)} \text{, with } x=\frac{c_\text{fcl}\, R}{R_\lambda}
\end{equation}
where $\tilde{f}(x)$ is the radial trend of a projected NFW (which we normalise at the scale radius), and $R_\lambda = (\lambda/100)^{0.2}\,h^{-1}$ Mpc. 
The parameters $(A_\text{fcl},\, B_\text{fcl}, c_\text{fcl})$ are the amplitude, the richness trend and the concentration of the cluster member contamination profile, respectively.  These parameters can be constrained well even in rather small samples of, e.g. SZE or X-ray selected clusters (Paulus et al., in prep). The concentration typically takes a low value, $c\approx2.5$, while the redshift trend is weak, $B_\text{fcl}\approx0.15$, and the amplitude $A_\text{fcl}\approx0.3$ represents $30\%$ contamination at the scale radius. Varying the parameters $(A_\text{fcl},\, B_\text{fcl}, c_\text{fcl})$ then accounts for the systematic uncertainties associated with the cluster member contamination.

In our application, we produce the synthetic shear profile for the $i$-th cluster by diluting the measured shear profile by the cluster member contamination,
\begin{equation}
\hat g_\text{t}^i(R |R_\text{mis}) = \bigg(1-f_\text{cl}(R|\lambda^i) \bigg) g_\text{t, meas}^i(R | R_\text{mis}).
\end{equation}

\subsubsection{Library creation}
Going through the steps outlined in the previous sections, we can simulate a realistic cluster catalog with an observable richness $\lambda$ and a mis-centering distribution for each halo. Furthermore, for each halo we also compute 3 shear profiles --  one for each cylindrical projection along each Cartesian axis -- for each of the mis-centering radii we consider. This shear library is created for a set of WL systematics and cluster parameters drawn from priors, which are presented in Section~\ref{sec:priors}.
Fig.~\ref{fig:shear_library} shows one realisation of such a shear library: in the left panel stacked in different mass bins, in the center showing the impact of different mis-centering radii, and in the right panel at different redshift. We present the shear profiles weighed by the square root of the area in units of degrees. The reported magnitude of the signal can thus be directly compared to the shape noise in the tangential shear measurement, which goes as $\sim 0.3 / \sqrt{ n_\epsilon }$, where $n_\epsilon$ is the source galaxy density in deg$^{-2}$. This weighting highlights the radial range in which the number of shapes and the strength of the density contrast provide the most signal, and it cancels the redshift trend introduced by the smaller solid angle subtended by a radial bin of fixed metric size.

\subsection{Shear profile model and measurement of $M_\text{WL}$}
\label{sec:shear_model}

\begin{figure*}
	\includegraphics[width=\textwidth]{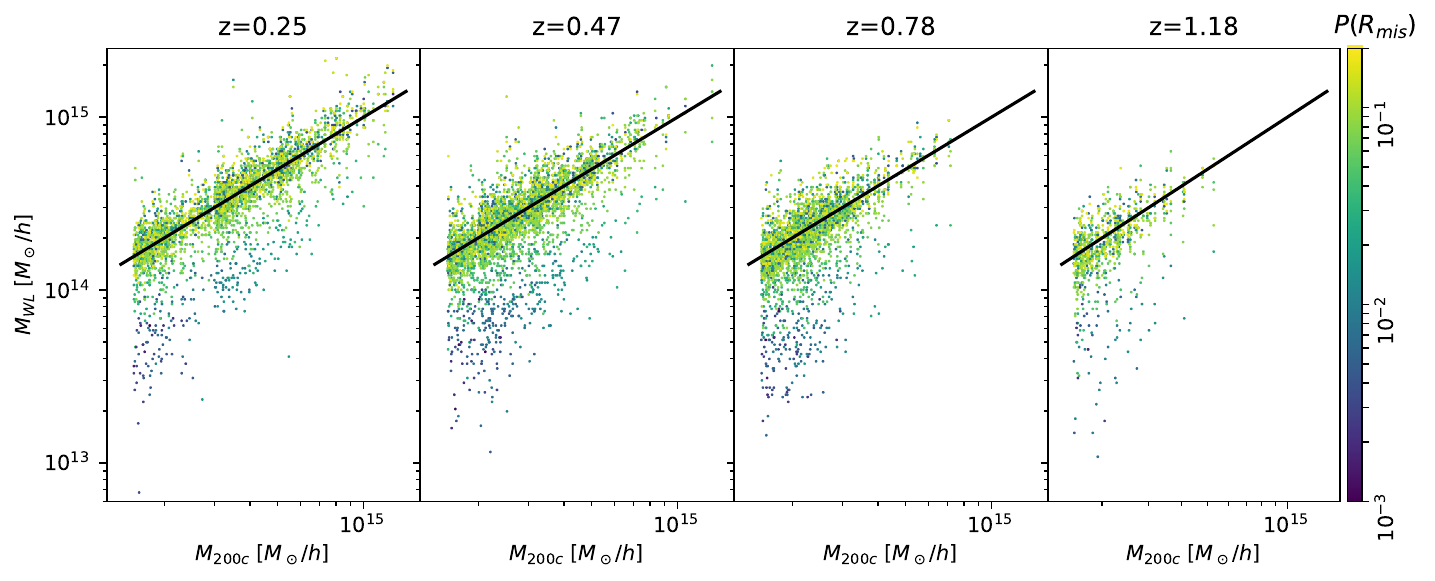}
	\vskip-0.10in
    \caption{For a single realisation of WL systematics we can extract the so called weak-lensing mass for each halo projection as a function of the mis-centering used in the extraction by assuming a shear profile model. We show the weak-lensing mass versus halo mass in the four different simulation snapshots. Color-coded is the probability weight, derived from the probability of the mis-centering. Strongly mis-centered simulated profiles have highly biased weak-lensing masses, but they are also very improbable. The black line shows the 1--1 relation.}\label{fig:mWL_vs_mhalo}
\end{figure*}

To capture the statistical and systematic uncertainties in the weak-lensing measurement, we measure the weak-lensing mass $M_\text{WL}$ and quantify its relationship to the true halo mass. The weak-lensing mass results from fitting
the simulated (or measured) shear profiles onto a shear profile model \citep{applegateetal14, schrabback18a, dietrich19, stern19}. As such, the weak-lensing mass is dependent on the shear profile model. 
This gives us the freedom to define a model which is simple to compute
at the expense of not being the most accurate on a cluster-by-cluster basis.
, but might not be the most accurate on a cluster by cluster basis. These inaccuracies lead to biases and scatter between the weak-lensing and halo mass. The ultimate goal of our method is a robust determination of this bias and scatter and their respective uncertainties.

In the following we discuss the shear profile model we use to compress the complex and realistic shear profiles simulated in our shear library into weak-lensing masses.
First, we extract the WL masses assuming a NFW model $\Sigma_\text{NFW}(R |M_\text{200c})$, and a constant concentration $c_\text{NFW}=3.5$, a typical concentration for massive halos \citep[e.g.][]{child18}. Fixing the concentration to a constant value 
simplifies the WL mass extraction in the cosmological analysis at the cost of increasing the WL bias and scatter.

Second, we use a simplified model for the mis-centered matter surface density
\begin{equation}
    \Sigma(R |M, R_\text{mis})= 
\begin{cases}
\begin{split}
    \Sigma_\text{NFW}(R_\text{mis}| M_\text{200c})  &\text{ for }R<R_\text{mis}&\\
    \Sigma_\text{NFW}(R |M_\text{200c}) &\text{ for }R>R_\text{mis}. &
\end{split}
\end{cases}
\end{equation}
This model is expected to lead to some inaccuracy in the radial ranges $R\sim R_\text{mis}$, which
lead to calibrateable biases
between $M_\text{WL}$ and the halo mass. However, our model simplifies the cosmological inference, 
by avoiding the explicit calculation of the azimuthal average 
\begin{equation}
\begin{split}
    \Sigma(R| & R_\text{mis}, M_\text{200c}) =  \\ 
    & = \frac{1}{2\pi} \int_0^{2\pi} \text{d}\theta \, \Sigma\bigg(\sqrt{ R^2+R_\text{mis}^2-2RR_\text{mis}\cos\theta}\bigg|M_\text{200c}\bigg). 
    \end{split}
\end{equation}

Third, instead of marginalising over the distribution of $R_\text{mis}$
we define one mis-centering radius for the extractions 
    \begin{equation}
        R_\text{mis}^{\text{extr},i} = R_{\lambda^i} \sqrt{\frac{\pi}{2}}\bigg(\bar\rho \bar\sigma_0 + (1-\bar\rho)\bar\sigma_1 \bigg),
    \end{equation}
that is the mean mis-centering radius for the mis-centering distribution evaluated at the mean mis-centering parameters $(\bar\rho, \bar\sigma_{0,1})$, evaluated for the measured richness $\lambda^i$. Other works use centered NFW profiles for the weak-lensing mass measurement, leading to less well behaved weak-lensing--halo mass distributions \citep{sommerip}.
Given the model for the matter surface density $\Sigma(R|R_\text{mis}^{\text{extr},i}, M)$, we can easily compute the density contrast $\Delta\Sigma(R|R_\text{mis}^{\text{extr},i}, M_\text{200c})$.

Fourth, we compute the lensing efficiency $\Sigma_\text{crit}^{-1}$ by setting $\alpha_\beta=0$ in Eq.~\ref{eq:beta}, i.e. just assuming the mean photo-z bias.

Fifth,
we compute the convergence $\kappa=\Sigma_\text{crit}^{-1} \Sigma(R|R_\text{mis}^{\text{extr},i}, M)$ and the shear $\gamma_\text{t} =\Sigma_\text{crit}^{-1} \Delta\Sigma(R|R_\text{mis}^{\text{extr},i}, M)$, as well as the reduced shear $g_\text{t}(R|R_\text{mis}^{\text{extr},i}, M_\text{200c}) = \gamma_\text{t} / (1-\kappa)$. Our final model for the measured reduced shear then takes account of the mean cluster member contamination profile $\bar f_\text{cl}(R|\lambda^i)$ and the mean multiplicative shear bias $m$, i.e.
\begin{equation}
    g_\text{t}^\text{mdl}(R|M_\text{200c}, \lambda^i) = (1+m)\bigg(1-\bar f_\text{cl}(R|\lambda^i)\bigg) g_\text{t}(R|R_\text{mis}^\text{extr}, M_\text{200c})
\end{equation}

And sixth, we measure the weak-lensing mass by minimizing the radial bin area--weighted difference between the simulated and model shear profiles. The weak-lensing mass for the $i$-th cluster, when taking the simulated profile mis-centered by $R_\text{mis}$, then is
\begin{equation}
    M_\text{WL}^i(R_\text{mis}) = \min_{M_\text{200c}} \sum_k A_k \bigg(g_\text{t}^\text{mdl}(R_k|M_\text{200c}, \lambda^i)-\hat g_\text{t}^i(R_k|R_\text{mis}) \bigg)^2,
\end{equation}
where $k$ runs over the radial bins, and $A_k$ is the area covered by that bin. The shape measurement variance in real measurements scales like the inverse of the bin area. \citet{sommerip}
explicitly show how this setup produces unbiased weak-lensing mass estimates independently on the amount of shape noise. This weight guarantees that we are weighting the different scale in the same way as they will be weighted in the real data. Note also that we added the uncorrelated LSS noise to the shear profile, rather than considering it a noise source in the $M_\text{WL}$ extraction. This implies that the uncorrelated LSS variance contributes to the WL scatter $\sigma_\text{WL}$, as it is a statistical noise source on the shear profile. This configuration ensurers that while extracting $M_\text{WL}$ we apply the correct relative weights to the different scales.

The choice of the innermost radial scale $R_\text{min}$ and outermost radial scale $R_\text{max}$ has a great impact on the accuracy of the weak-lensing mass extraction. As a baseline we choose $R_\text{min}=0.5\,h^{-1}$ Mpc, and $R_\text{max} = 3.2 (1+z_\text{cl})^{-1}\,h^{-1}$ Mpc. We will explore the impact of varying the inner fitting radius and also argue for the redshift dependence of the outer fitting radius below (cf. Section~\ref{sec:inner_fitting_radius}, and Section~\ref{sec:outer_radius}).

This procedure provides us with 3 $M_\text{WL}^i(R_\text{mis})$ for each halo and for each $R_\text{mis}$ thanks to the 3 projection axes. It is noteworthy here that thanks to the mis-centering distribution $P(R_\text{mis}|\lambda^i)$ we also know how probable each mis-centering radius $R_\text{mis}$ is. 

\begin{figure}
	\includegraphics[width=\columnwidth]{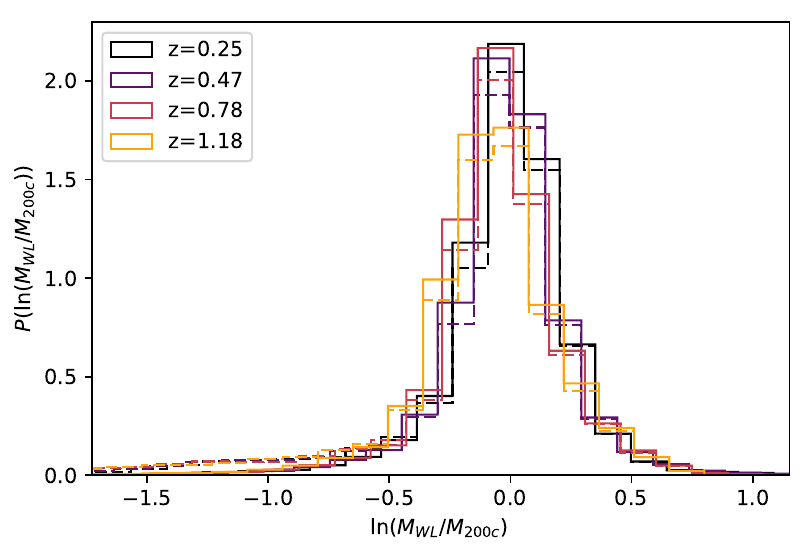}
	\vskip-0.10in
    \caption{Distribution of weak lensing mass $M_\text{WL}$ around the halo mass for different redshifts (dashed line without the weighting by the mis-centering distribution, full line with the weighting by the mis-centering distribution) for a single realisation of the shear library. Applying the weighting by the mis-centering distribution down-weights the lowest extracted weak lensing masses, which correspond to the largest mis-centering radii. The resulting distribution is symmetric, although fat-tailed. We fit this distribution with a log-normal whose mean and variance can vary with halo mass and redshift, postponing the investigation of the tails to future work.}\label{fig:mass_resiudals}
\end{figure}

Fig.~\ref{fig:mWL_vs_mhalo} shows the scatter plot between the weak-lensing mass and the halo mass in the five different snapshots we analysed. Color coded is the associated probability $P(R_\text{mis}|\lambda^i)$. Clearly, some weak-lensing masses underestimate the halo mass by more than an order of magnitude. These are associated to large mis-centering radii $R_\text{mis}$. However, such large mis-centering radii are also highly improbable. Indeed, for the most probable mis-centering radii, the weak-lensing mass is comparable to the halo mass, albeit with some scatter. Statistically speaking the highly biased weak-lensing masses due to strong mis-centering are not very relevant, as can be seen in Fig.~\ref{fig:mass_resiudals}, which shows the probability weighted distribution of the ratio between weak-lensing mass and halo mass (solid line) in comparison to the raw distribution (dashed line). The weighting suppresses the left tail of the distribution. The weak-lensing masses extracted for probable mis-centering radii scatter around a ratio of $\sim$1, indicating that our simple model for the shear profile provides an adequate fit to the data.

Following this approach, we are able to compress the complex shear profiles in our realistic shear profile library into weak lensing masses. The model defining the weak-lensing mass does not
explicitly
take account of the different statistical noise sources impacting the simulated shear profile. These statistic noise sources thus manifest as scatter between the weak-lensing and the halo mass. Similarly, our model does not explicitly account for systematic uncertainties in the halo catalog and shear profiles. Instead, fitting a relation for the weak-lensing--halo mass parameters with different realisations of weak-lensing systematics allows us to propagate their uncertainty. 

\subsection{Relation between weak-lensing and halo mass}\label{sec:shearmodel&mWL}

\subsubsection{Fitting}\label{sec:fitting_WL-halo_rel}
In the previous section, we established the weak-lensing mass such that it closely tracks the halo mass with some scatter and some (small) amount of bias. 
For the mean relation, we allow for a complex redshift trend $b_\text{WL}(z)$ and a power-law trend in mass, i.e.
\begin{equation}\label{eq:meanMWLgivenM}
    \Big\langle \ln \Big(\frac{M_\text{WL}}{M_0}\Big)\Big\rangle = 
    b(z)+ b_\text{M} \ln \Big(\frac{M_\text{200c}}{M_0}\Big),
\end{equation}
with the pivot mass $M_0=2\times 10^{14}\,h^{-1}$ M$_\odot$.
For the redshift evolution of the bias $b(z)$, we fit an independent bias parameter $b_{z_i}$ for each simulation snapshot at $z_i\in(0.25,\,0.47,\,0.78,\,1.18)$. 
For the scatter, we set up a log-normal model
\begin{equation}\label{eq:PlnMWLgM}
    P(\ln M_\text{WL}|M_\text{200c},z) = \mathcal{N}(\ln M_\text{WL}| \langle \ln M_\text{WL}\rangle, \sigma_\text{WL}^2)
\end{equation}
and we allow the variance to vary with mass and redshift
\begin{equation}\label{eq:varMWLgivenM}
    \ln \sigma_\text{WL}^2 = s(z) + s_\text{M}\ln \Big(\frac{M_\text{200c}}{M_0}\Big).
\end{equation}
 For the redshift evolution of the bias $s(z)$, we fit an independent bias parameter $s_{z_i}$ for each simulation snapshot at $z_i\in(0.25,\,0.47,\,0.78,\,1.18)$. 

To fit for the parameters of the weak-lensing mass -- halo mass relation we maximise the likelihood
\begin{equation}\label{eq:likeWLmass-mass}
    \ln L = \sum_i \ln \sum_{R_\text{mis}} P(R_\text{mis}|\lambda^i)P\bigg(\ln M_\text{WL}^i(R_\text{mis})\bigg|M^i,z^i\bigg),
\end{equation}
which ensures that the weak-lensing masses extracted from strongly mis-centered profiles are weighted by the mis-centering probability.
This approach implicitly marginalises over the mis-centering distribution, providing significant time savings in the cosmological pipeline.

We maximise the likelihood of the weak-lensing mass -- halo mass relation given in Eq.~\ref{eq:likeWLmass-mass}, resulting in a best-fit estimate of the parameters of the weak-lensing mass--halo mass relation $(b_{z_i},\,b_\text{M},\,s_0,\,s_\text{M},\,s^\prime)$, and in a covariance matrix $C_\text{sim}$ that describes the measurement uncertainties on these parameters that arise from the finite halo sample.

The scatter between weak-lensing mass and halo mass that we characterise here is due to several sources: 
variations in orientation and morphology of the cluster halos (i.e, ellipticity, concentration, and substructure due to the diversity of dynamical states reflected in the simulations),
correlated structure around the clusters,
uncorrelated large-scale structure, 
and the variance stemming from the range of possible mis-centering 
errors in choosing the cluster center. 
Our simple model allows us to 
marginalize over all these effects, producing posterior distributions in the weak-lensing mass to halo mass relation parameters that can then be incorporated within a weak-lensing cluster mass calibration analysis to account for systematic effects and the remaining uncertainties on those effects.

\subsubsection{Accuracy of the weak-lensing mass extraction}\label{sce:accuracy_WLhalo_rel}
The accuracy of the weak-lensing mass extracted using the described mapping are reflected in the statistical and systematic uncertainties at play in the approach described above.
The 
uncertainties on the parameters $(b_{z_i},\,b_\text{M},\,s_{z_i},\,s_\text{M})$ of the mapping between weak-lensing and halo mass 
stems in part from the finite number of halos for which we extracted matter profiles from the simulations. We shall denote this uncertainty $C_\text{sim}$ and note that these are ``statistical'' in nature, in that they can be reduced by enlarging the profile library.

Another source of uncertainty comes from the systematic uncertainties in the adopted parametrizations of the weak-lensing catalog, photometric source galaxy catalog and cluster properties.  These include the parameters that govern the halo catalog, i.e. the parameters of the observable--mass scaling relation $(A_\lambda,\, B_\lambda,\, C_\lambda, \ln\sigma_\lambda)$ (cf. Section~\ref{sec:obs_mass}), and the parameters of the mis-centering distribution $(\rho,\, \sigma_{0},\, \sigma_{1})$ (cf. Section~\ref{sec:mis_centering}), and the parameters of the weak-lensing systematics, i.e. the modulation of the photometric redshift uncertainties $\alpha_\beta$ (cf. Section~\ref{sec:photo-z_bias}), linear and non-linear shape measurement uncertainties $(\alpha_\gamma,\, \alpha_\text{NL})$ (cf. Section~\ref{sec:mult_shear_bias}), and the cluster member contamination $(A_\text{fcl},\, B_\text{fcl}, c_\text{fcl})$ (cf. Section~\ref{sec:member_cont}). By varying these parameters, we can create different realisations of shear libraries that sample the range of systematic uncertainty on the measured shear. 
To estimate the impact of these uncertainties, we draw from the best-fitting parameters and their uncertainties to create $\sim 1000$ different realisations of the parameters, following the prior listed below (cf. Section~\ref{sec:priors}). We then use each realization to create a new shear profile library, to extract the weak-lensing masses through our fitting process and to then fit the weak-lensing to halo mass relation. This results in a estimate of weak-lensing bias and scatter for each realisation of systematics. These result are then combined to build a joint posterior for the parameters of the weak-lensing to halo mass relation parameters $(b_{z_i},\,b_\text{M},\,s_{z_i},\,s_\text{M})$ that reflects the uncertainties from the different systematics.

This posterior completely charaterises the impact of modelled weak-lensing systematics on the relation between weak-lensing and halo mass. The width of the posterior on the weak-lensing halo mass relation parameters can then be used for cosmological analyses. The uncertainty induced by varying the weak-lensing systematics is larger than the noise due to the limited number of halos for which matter profiles were extracted from the simulations, $\sqrt{\text{diag}(C_\text{sim})}$.
As systematic uncertainties on the weak-lensing, photometric redshift and halo related biases will be better characterized in the future, it will be necessary to scale up the size of the shear profile library extracted from the simulations.

\subsubsection{Priors on weak-lensing systematics}\label{sec:priors}

Here we specify the ranges within which the weak-lensing systematics are allowed to vary. We caution the reader that the calibration of the weak-lensing mass extraction presented in the following is valid only for the specific shear profile model described in Section~\ref{sec:shear_model}, and the systematics treatment described in this paper. Application to any other scenario requires a dedicated analysis, which would follow the same procedure outlined here.

For the parameters of the richness--mass relation, we follow the recent analysis by \citet{bleem19} $A_\lambda\sim\mathcal{N}(78.5;8.2^2),\,B_\lambda\sim\mathcal{N}(1.02;0.08^2),\,C_\lambda\sim\mathcal{N}(0.29;0.27^2),\, \ln\sigma_\lambda\sim\mathcal{N}(\ln 0.23; 0.16^2)$. For the parameters of the mis-centering distribution, we follow \citet{saro15} for the central values and the uncertainties for $\rho\sim\mathcal{N}(0.63;0.06^2),\,\sigma_0\sim\mathcal{N}(0.07;0.02^2),\,\sigma_1\sim\mathcal{N}(0.25;0.07^2)$, but we reduce the uncertainty by half.

The parameters $\alpha_{\beta, m}$ are introduced with the purpose of marginalising over the multiplicative shear bias uncertainty (cf. Section~\ref{sec:mult_shear_bias}) and the photo-z bias uncertainty (cf. Section~\ref{sec:photo-z_bias}). This is accomplished by sampling them $\alpha_{\beta, m}\sim\mathcal{N}(0;1)$. For the strength of the non-linear shear response, we rely on the argument given in \citet{sheldon17}. For the DES Y1 data, they claim that $\alpha_\text{NL}=0.6$ with 40\% uncertainty, i.e. $\ln \alpha_\text{NL} \sim \mathcal{N}(\ln 0.6; 0.4^2)$, would not bias the cosmic shear results. 
 
For the cluster member contamination, we assume $\ln A_\text{fcl}\sim\mathcal{N}(\ln 0.3; 0.1^2),\,B_\text{fcl}\sim\mathcal{N}(0.15;0.1^2),\,c_\text{fcl}\sim\mathcal{N}(2.5;0.1^2)$. This comes down to a $\sim$30$\%$ contamination at the scale radius, with a weak richness trend. The radial trend is an NFW whose concentration is comparable to the concentration of blue galaxies in clusters \citep[see][]{hennig17}, as blue galaxies are plausibly expected to contribute more to the cluster member contamination due to their lower quality photo-z's.

\subsubsection{Tracking the sources of uncertainty}\label{sec:tracking}
Consider a multivariate Gaussian distribution with means 0 and a covariance $C_{i,j} = \sigma_i\sigma_j \rho_{i,j}$, with $\rho_{i,j}=1$ if $i=j$. The variance on $x_k$ is $\text{Var}[x_k]=\sigma_k^2$, and the variance on $x_k$ conditional on $x_l$ is $\text{Var}[x_k | x_l] = \sigma^2_k (1-\rho_{k,l}^2)$.
This can be interpreted as ``knowing $x_l$ reduces the variance on $x_k$ by $\rho_{k,l}^2$ percent''. Conversely, this means that the uncertainty on $x_{l}$ propagates to an increase in the variance of $x_k$ equal to $\sigma^2_k \rho_{k,l}^2$. Reporting the squared correlation coefficients between the weak-lensing--halo mass parameters $(b_{z_i},\,b_\text{M},\,s_{z_i},\,s_\text{M})$ and the weak-lensing systematics parameters $(A_\lambda,\, B_\lambda,\, C_\lambda, \ln\sigma_\lambda, \rho,\, \sigma_{0},\, \sigma_{1},\, \sigma_{0}^\prime,\, \sigma_{1}^\prime\alpha_\beta,\,\alpha_\gamma,\, \alpha_\text{NL},\, A_\text{fcl},\, B_\text{fcl}, c_\text{fcl})$ thus provides a means of quantifying the impact of each systematics parameter on each of the weak-lensing to halo mass parameters.

We estimate the squared correlation coefficient between the parameters $x$ and $y$ from the sample of $\{ x_i,\,y_i\}$ as
\begin{equation}
    \rho^2_{x,y} = \frac{N^{-1} \sum_i (x_i -\bar x) (y_i -\bar y)}{\sqrt{\text{Var}[x]\,\text{Var}[y]}}, 
\end{equation}
with the variance on $x$ estimated as
\begin{equation}
    \text{Var}[x] = N^{-1} \sum_i (x_i -\bar x)^2.
\end{equation}
Here $N$ is the  number of samples used, and $\bar x$, and $\bar y$ are the means of $x$, and $y$, respectively.

\subsubsection{Skewness and kurtosis}
To test whether the assumed log-normal model in Eq.~\ref{eq:PlnMWLgM} is adequate, we compute the skewness and kurtosis of the residuals between the extracted weak-lensing mass $\ln M_\text{WL}^i(R_\text{mis})$ and the predicted weak-lensing mass $\langle \ln M_\text{WL}\rangle(M_\text{200c}^i, z^i)$. For simplicity we define the residual $\Delta\ln M_\text{WL}^i(R_\text{mis})=\ln M_\text{WL}^i(R_\text{mis})-\Big\langle \ln M_\text{WL}\Big\rangle(M_\text{200c}^i, z^i)$. We also need to consider that each data point has a weight $w(R_\text{mis},i) = P(R_\text{mis}|\lambda^i)$.

The skewness of a weighted sample is given by
\begin{equation}
    \gamma_\text{skw} = \frac{1}{W} \sum_{R_\text{mis},i} w(R_\text{mis},i) \bigg(\Delta\ln M_\text{WL}^i(R_\text{mis}) \bigg)^3,
\end{equation}
while the kurtosis is given by 
\begin{equation}
    \gamma_\text{krt} = \frac{1}{W} \sum_{R_\text{mis},i} w(R_\text{mis},i) \bigg(\Delta\ln M_\text{WL}^i(R_\text{mis}) \bigg)^4,
\end{equation}
with $W=\sum_{R_\text{mis},i} w(R_\text{mis},i)$.

We estimate the statistical error on the skewness and kurtosis via bootstrapping. We also study the impact of the uncertainty on the weak-lensing systematics on skewness and kurtosis, by tracking them for each realisation of shear library we study.  We then consider deviations of the skewness and kurtosis from the Gaussian case ($\gamma_\text{skn}=0$ and $\gamma_\text{krt}=3$) as indicators if our assumed log-normal scatter model is adequate.

\section{Results and Discussion}\label{sec:results}

In the following we will present the results from the calibration of the weak-lensing bias and scatter of our generic weak-lensing survey (Section~\ref{sec:calib_gen_survey}). We then discuss the choice of the inner and outer fitting radius (Section~\ref{sec:inner_fitting_radius}, and Section~\ref{sec:outer_radius}, respectively), and the impact of further systematics that are harder to model to date (Section~\ref{sec:further_sys}). Finally, we put our 
 framework for dealing with systematics related to hydrodynamical effects
in context with other works (Section~\ref{sec:hydro_approaches}).

\subsection{Calibration for generic catalog and weak-lensing survey}\label{sec:calib_gen_survey}

\begin{figure*}
	\includegraphics[width=\textwidth]{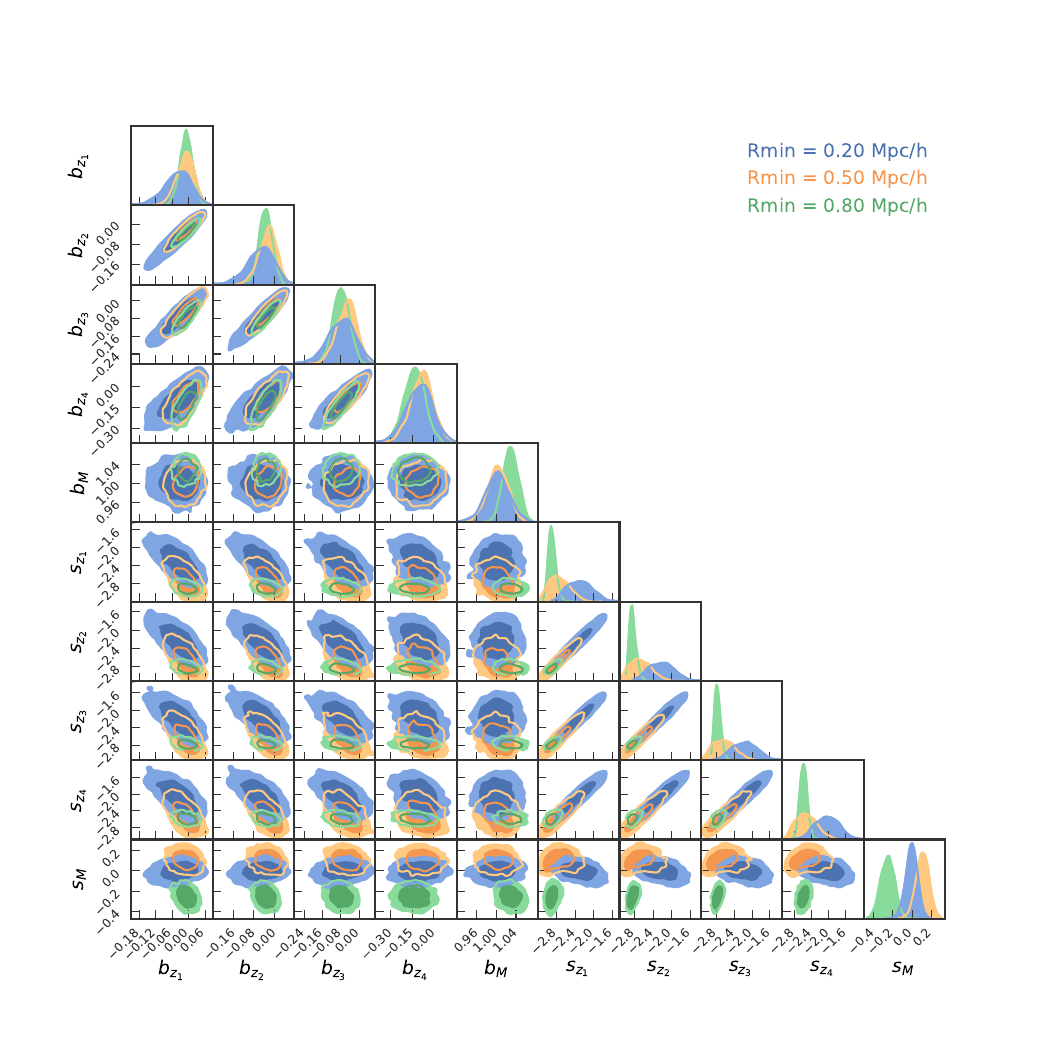}
	\vskip-0.15in
    \caption{Posterior on the weak-lensing to halo mass relation parameters (equations~\ref{eq:meanMWLgivenM} to \ref{eq:varMWLgivenM}) extracted from the analysis of $\sim$1000 different realisations of shear libraries with different weak-lensing and cluster catalog systematics. $b_{z_i}$ are the biases between the weak-lensing and the halo mass for each simulation snapshot at $z_i\in(0.25,\,0.47,\,0.78,\,1.18)$, $b_\text{M}$ is the  mass trend of these biases, while $s_{z_i}$ are the natural logarithms of the variance of the weak-lensing mass at given halo mass, and $s_\text{M}$ is its mass trend. The posterior displays strong degeneracies among different parameters. It encapsulates the systematic uncertainty on the mapping between weak-lensing and halo mass, and can be used as a prior for cosmological analyses. We present the posterior for different choices of radial fitting range. Generally, using smaller scales for the fit increases the uncertainty on the weak-lensing to halo mass scaling parameters, because central cluster regions are more affected by, e.g.  mis-centering and cluster contamination.}\label{fig:contour_standard}
\end{figure*}

\begin{figure*}
	\includegraphics[width=0.95\columnwidth]{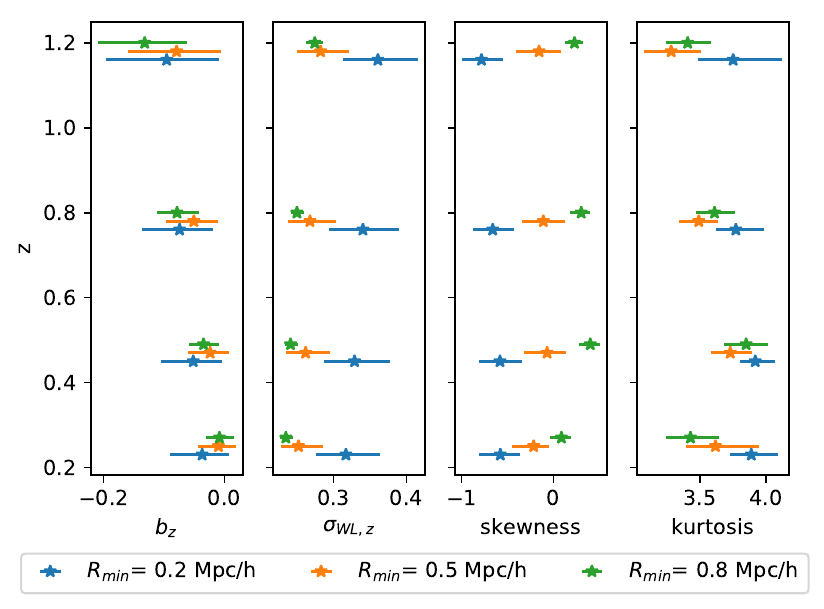}
	\includegraphics[width=1.05\columnwidth]{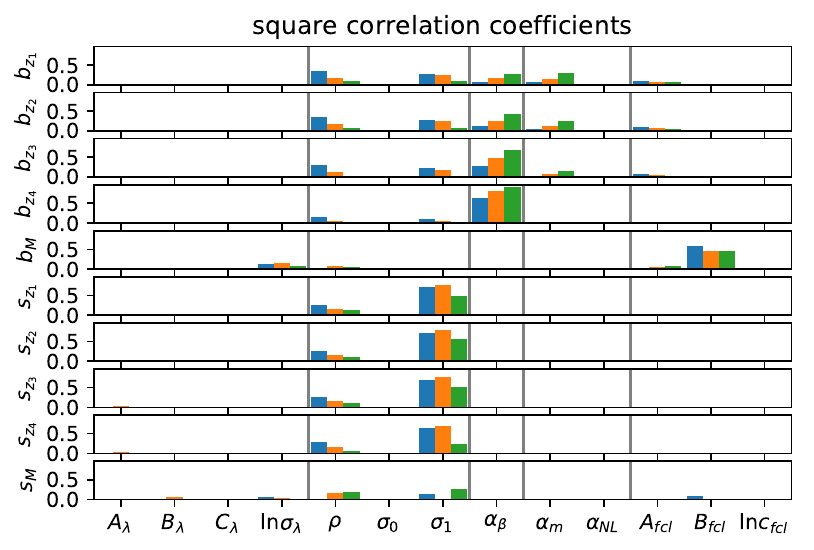}
	\vskip-0.10in
    \caption{\textit{Left panel:} Weak-lensing bias, weak-lensing intrinsic scatter, skewness and kurtosis (0 and 3, respectively, for a perfectly log-normal model) of the weak lensing mass -- halo mass distribution. \textit{Right panel:} Squared correlation coefficients between the weak-lensing--halo mass parameters and the weak-lensing and cluster catalog systematics parameters (i.e. the paramaters of the observable mass scaling relation $(A_\lambda,\, B_\lambda,\, C_\lambda, \ln\sigma)$, of the mis-centering distribution $(\rho,\, \sigma_{0},\, \sigma_{1})$, the modulation of the photometric redshift uncertainties $\alpha_\beta$, linear and non-linear shape measurement uncertainties $(\alpha_\gamma,\, \alpha_\text{NL})$, and the cluster member contamination $(A_\text{fcl},\, B_\text{fcl}, c_\text{fcl})$).  These correlations can be interpreted as the impact of a given systematic on weak-lensing--halo mass parameters. The colors stand for different choices of fitting range. Clearly, the mis-centering and cluster member contamination contribute more to the systematics budget if smaller radial scales are considered. Furthermore, the photo-z bias has as a stronger impact at high redshift, where it is less well constrained.}\label{fig:standard_corrs}
\end{figure*}

\begin{table}
	\centering
	\caption{Medians with difference to the 16th and 84th percentile of the parameters of the weak-lensing to halo mass relation obtained from several realisations of generic surveys assuming different inner fitting radii $R_\text{min}$.}\label{tab:results}
	\begin{tabular}{llll}
	\hline
	$R_\text{min}$   & 0.2 $h^{-1}$Mpc  & 0.5 $h^{-1}$Mpc & 0.8 $h^{-1}$Mpc  \\
\hline
$b_{z_1}$ & -0.037$^{+0.044}_{-0.053}$ & -0.010$^{+0.029}_{-0.033}$ & -0.008$^{+0.024}_{-0.023}$ \\[3pt]
$b_{z_2}$ & -0.052$^{+0.048}_{-0.054}$ & -0.024$^{+0.032}_{-0.037}$ & -0.035$^{+0.025}_{-0.025}$ \\[3pt]
$b_{z_3}$ & -0.074$^{+0.056}_{-0.062}$ & -0.051$^{+0.040}_{-0.047}$ & -0.079$^{+0.036}_{-0.034}$ \\[3pt]
$b_{z_4}$ & -0.096$^{+0.087}_{-0.100}$ & -0.079$^{+0.074}_{-0.081}$ & -0.132$^{+0.070}_{-0.077}$ \\[3pt]
$b_{M}$ & 1.002$^{+0.024}_{-0.024}$ & 1.003$^{+0.021}_{-0.021}$ & 1.029$^{+0.016}_{-0.016}$ \\[3pt]
$s_{z_1}$ & -2.298$^{+0.279}_{-0.277}$ & -2.758$^{+0.248}_{-0.206}$ & -2.901$^{+0.084}_{-0.069}$ \\[3pt]
$s_{z_2}$ & -2.224$^{+0.277}_{-0.269}$ & -2.682$^{+0.240}_{-0.222}$ & -2.847$^{+0.083}_{-0.059}$ \\[3pt]
$s_{z_3}$ & -2.156$^{+0.272}_{-0.291}$ & -2.637$^{+0.253}_{-0.241}$ & -2.776$^{+0.080}_{-0.061}$ \\[3pt]
$s_{z_4}$ & -2.038$^{+0.284}_{-0.282}$ & -2.531$^{+0.258}_{-0.243}$ & -2.588$^{+0.084}_{-0.087}$ \\[3pt]
$s_{M}$ & -0.005$^{+0.057}_{-0.061}$ & 0.110$^{+0.065}_{-0.068}$ & -0.253$^{+0.069}_{-0.076}$ \\[3pt]
$\gamma_\text{skw,1}$ & -0.578$^{+0.217}_{-0.229}$ & -0.215$^{+0.166}_{-0.231}$ & 0.088$^{+0.108}_{-0.129}$ \\[3pt]
$\gamma_\text{skw,2}$ & -0.582$^{+0.239}_{-0.229}$ & -0.069$^{+0.205}_{-0.252}$ & 0.402$^{+0.104}_{-0.121}$ \\[3pt]
$\gamma_\text{skw,3}$ & -0.661$^{+0.236}_{-0.215}$ & -0.109$^{+0.232}_{-0.231}$ & 0.303$^{+0.091}_{-0.123}$ \\[3pt]
$\gamma_\text{skw,4}$ & -0.783$^{+0.230}_{-0.213}$ & -0.156$^{+0.236}_{-0.247}$ & 0.230$^{+0.097}_{-0.107}$ \\[3pt]
$\gamma_\text{krt,1}$ & 3.888$^{+0.203}_{-0.159}$ & 3.616$^{+0.332}_{-0.225}$ & 3.426$^{+0.220}_{-0.190}$ \\[3pt]
$\gamma_\text{krt,2}$ & 3.918$^{+0.149}_{-0.119}$ & 3.729$^{+0.169}_{-0.148}$ & 3.849$^{+0.168}_{-0.167}$ \\[3pt]
$\gamma_\text{krt,3}$ & 3.771$^{+0.216}_{-0.152}$ & 3.489$^{+0.148}_{-0.150}$ & 3.609$^{+0.157}_{-0.139}$ \\[3pt]
$\gamma_\text{krt,4}$ & 3.750$^{+0.370}_{-0.269}$ & 3.280$^{+0.223}_{-0.207}$ & 3.405$^{+0.179}_{-0.166}$ \\[3pt]
\hline
	\end{tabular}
\end{table}

We analyse the generic survey described above with three different settings for the minimal fitting radius: $R_\text{min}=0.2,~0.5,~0.8\,h^{-1} \,\mathrm{Mpc}$. This results in the three posterior distributions whose 1-d  and 2-d marginals are shown in Fig.~\ref{fig:contour_standard}. The 1-d median parameter values and their difference with the 16th and 84th percentile values are reported in Tab.~\ref{tab:results} and shown in Fig.~\ref{fig:standard_corrs}, left panel. Our weak-lensing mass is biased low by 0 to 13\%, depending mainly on redshift. We find a mass trend close to linear.  

The exact numerical values of the biases are subject to the details of the systematics implemented in the creation of the generic survey (cf. Section~\ref{sec:shear_lib}) and the specific model choices used to measure the weak-lensing mass (cf. Section~\ref{sec:shear_model}).
In particular, the biases could be reduced by adjusting
our shear model
to better represent the simulated shear profiles, but as long as the bias remains relatively small it is the uncertainty on this mean bias rather than the bias itself that serves as a limitation on the ensuing weak-lensing mass calibration.
  
The intrinsic scatter about the weak-lensing to halo mass relation is between 0.2 and 0.3 in the natural logarithm of the weak-lensing mass. It depends mainly on the inner fitting radius; for smaller inner fitting radii the intrinsic scatter is larger. The intrinsic scatter shows no clear trend with mass, but the uncertainties are large. A shear model that minimizes this variance would be crucial to any weak-lensing mass calibration, because a smaller variance effectively increases the mass information coming from each weak-lensing mass measurement.

For the smallest fitting radius considered  ($R_\text{min}=0.2\,h^{-1}\,\mathrm{Mpc}$) the distribution of weak-lensing masses given halo mass is negatively skewed, while it shows no skewness for $R_\text{min}=0.5\,h^{-1}\,\mathrm{Mpc}$ and a positive skewness for $R_\text{min}=0.8\,h^{-1}\,\mathrm{Mpc}$. As shown in Appendix~\ref{sec:app_A}, the weak-lensing mass in absence of mis-centering is also positively skewed, which matches the large minimal fitting radius case (in which the impact of mis-centering is minimal). As can be seen in Fig.~\ref{fig:mWL_vs_mhalo}, strong mis-centering introduces weak-lensing masses that scatter low very strongly. When moving to smaller fitting radii, the impact of mis-centering increases. This manifests itself in the negative skewness for the $R_\text{min}=0.2\,h^{-1}\,\mathrm{Mpc}$ case. For the current setting, the two effects seem to cancel in the $R_\text{min}=0.5\,h^{-1}\,\mathrm{Mpc}$ case, leading to no detectable skewness, but this might not be the case for different mis-centering distributions.  

Note also that the kurtosis is larger than 3 (the Gaussian case) for all settings studied here. That means that distribution of weak-lensing masses given halo mass is fat-tailed. The impact of these tails will be investigated in future work.

Perhaps the most important result is the uncertainty on the different parameters of the weak-lensing to halo mass relation. They reflect uncertainties on the weak-lensing systematics parameters, and these ultimately limit the accuracy and precision with which a weak-lensing mass calibration can be done.
We find that for a inner fitting radius of $R_\text{min}=$ 0.2 (0.5, 0.8) $h^{-1}\,\mathrm{Mpc}$ the uncertainty on the bias parameters increases with redshift from 4.8\% (3.1\%,  2.3\%) at z=0.25 to 9.4\% (7.8\%, 7.4\%) at z=1.18. The same trend can also be seen for the mass trend of the bias and the intrinsic variances.  This quantitatively confirms an important aspect of the treatment of systematics in cluster weak-lensing studies: the closer one fits to the cluster center, the larger the uncertainties on the weak-lensing to halo mass relation parameters and therefore the larger the systematics budget in a weak-lensing mass calibration analysis.

Finally, we explore the correlation between the parameters of the weak-lensing to halo mass relation and the weak-lensing systematics parameters. This allows us to gauge which systematic effects are most important in determining the mapping from weak-lensing to halo mass (cf. Section ~\ref{sec:tracking}).

The squared correlation coefficients are shown in Fig.~\ref{fig:standard_corrs} (right), where the weak-lensing to halo mass parameters are on the vertical axis and the weak-lensing systematics parameters are organized by function into five groups on the horizontal axis.  From left to right these are the parameters from the $\lambda$-mass relation, mis-centering, photometric redshifts, shear bias and finally cluster member contamination. 

By inspecting the first 4 rows, we can see that at larger redshifts the biases $b_{z}$ are impacted by the photometric redshift uncertainty ($\alpha_\beta$ column) the most. We can also see that the amplitude of the cluster member contamination ($A_\text{fcl}$ column), the multiplicative shear bias ($\alpha_m$ column) and the weight and amplitude of the larger mis-centring distribution component ($\rho$ and $\sigma_1$ columns) correlate with the biases (and thus contribute to their systematic uncertainty). We can also see that the contribution from mis-centering and cluster member contamination diminishes for larger inner fitting radii, while the impact of multiplicative shear bias increases. 

Noticeably, we find no correlation with the parameters governing the richness--mass relation. 
Remember that we introduced these parameters because
some parts of our shear model depend on the measured richness. The accuracy of our weak-lensing mass calibration is thus unaffected by changes in the parameters of the richness mass relation. It is also worth pointing out that the uncertainty on the non-linear shear response, albeit very large, plays no role in the overall systematics budget. Other cluster weak-lensing systematics have larger impacts. This is reassuring, as the uncertainty on the non-linear shear response was chosen in accordance with the limits required by cosmic shear experiments. 

\subsection{Inner fitting radius}\label{sec:inner_fitting_radius}

\begin{figure}
	\includegraphics[width=\columnwidth]{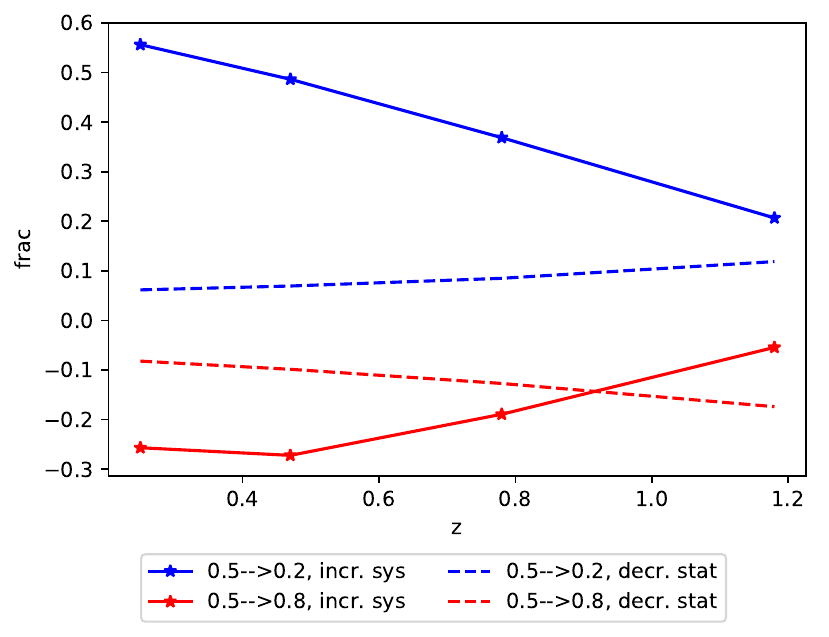}
	\vskip-0.10in
    \caption{Fractional increase in the systematic uncertainty on the bias parameters (solid line) when changing the inner fitting radius, compared to fractional change in statistical power (dashed line). Changing the inner fitting radius from $R_\text{min}=0.5\,h^{-1}\,\mathrm{Mpc}$ to $R_\text{min}=0.2\,h^{-1}\,\mathrm{Mpc}$ results in a up to 50\% increase in the systematics floor but only a $\sim$10\% gain in statistical precision.  The optimal choice depends on the relative scale of the statistical and systematic errors, which in turn
    depends on the source galaxy and cluster number and redshift distributions. 
    }\label{fig:going_in?}
\end{figure}

We presented above the impact of the inner fitting radius choice on the calibration of the weak-lensing to halo mass relation. We found that using a larger inner fitting radius reduces the systematic uncertainty on the weak-lensing to halo mass bias parameters. In the context of a cosmological study, this means that in an experiment similar to the generic Stage III case adopted here, the systematic floor on the weak-lensing mass calibration is reduced by avoiding the inner core region of the cluster.

Using a larger fitting radius will however result in the exclusion of some shear data, reducing the precision of the measurement. This effect scales more quickly than the mere change in number of background galaxies, because the shear signal is stronger in the inner regions. Fig.~\ref{fig:shear_library} shows the shear signal-to-noise (shear signal scaled by square root of area on the sky) profiles for clusters at a range of redshifts and masses. In the range between 0.2-0.8 $h^{-1}$ Mpc the signal-to-noise is quite constant with radius (which means that the shear signal falls off approximately linearly with radius).  This suggests that the loss of signal from excluding the central core is not as severe as one might first think.

In Fig.~\ref{fig:going_in?} we contrast the increase in systematic uncertainty (ratio between the uncertainties of the biases) with the decrease of the statistical signal-to-noise (ratio of the total signal-to-noises in the respective fitting ranges). We can see that the change in signal-to-noise is 10-20\% when going from 0.5 to 0.8~$h^{-1}$Mpc or from 0.5 to 0.2~$h^{-1}$Mpc. The change is slightly larger at higher redshifts because our outer fitting radius gets smaller with larger redshift (see discussion in next section). Note that this estimate is independent of the number of clusters and the background source density (which would be the same no matter the choice of inner fitting radius). 

This modest change in statistical constraining power needs to be contrasted with the larger change in systematic uncertainty. Going further in ($R_\text{min}=0.2\,h^{-1}\,\mathrm{Mpc}$ instead of 0.5~$h^{-1}$Mpc) increases the systematic uncertainty by 20--50\%. Staying further out ($R_\text{min}=0.8$~$h^{-1}$Mpc instead of 0.5~$h^{-1}$Mpc) reduces the systematic uncertainty by 10--30\%. Determining which fitting range is optimal depends also on the relative scale of the systematic and statistical noise, which for a given set of weak-lensing systematics inputs depends on the statistical constraining power of the weak-lensing calibration. This follows from the number of source galaxies and clusters and their respective redshift distributions. However, these estimates show that choosing a smaller fitting radius provides a benefit in constraining power that is fractionally smaller than its fractional cost in systematic accuracy. 
On the other hand, choosing a larger inner fitting radius roughly offsets gains and losses in the fractional precision and accuracy.

\subsection{Outer fitting radius}\label{sec:outer_radius}

\begin{figure*}
	\includegraphics[width=\textwidth]{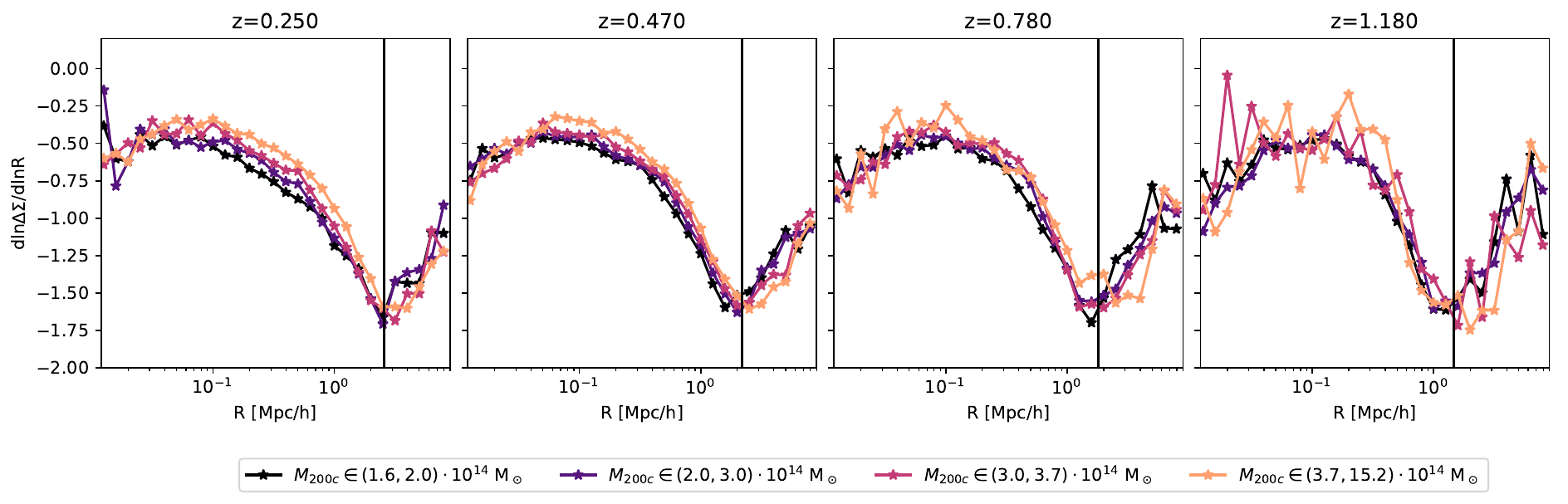}
	\vskip-0.10in
    \caption{Logarithmic slope of the density contrast profiles stacked in mass bins for different redshifts. The vertical line indicates the upper limit of our fitting range. It clearly limits the mass fit to the range dominated by the 1-halo term and avoids fitting the 2-halo term, which can be recognized by the increase of the logarithmic slope, i.e. a flattening of the profile. In order to avoid the  2-halo regime, with its complex dependence of cosmology and its yet unknown sensitivity to hydrodynamical modelling effect, we propose to measure the halo mass within $3.2\,(1+z)^{-1}\,h^{-1}\,\mathrm{Mpc}$.}\label{fig:2halo}
\end{figure*}

The choice of the outer fitting radius is motivated by the attempt to avoid the so-called ``2-halo'' term. While the halo matter profile is the dominant source of shear on small scales, at larger scales the shear signal around clusters is dominated by surrounding halos. Indeed, halos are biased tracers of the matter distribution
, and -- to first approximation -- the two-halo term can be computed from the matter power spectrum and the halo bias.  Because the halo bias is mass dependent, a bias measurement provides mass information.  But the mass dependence of the bias is weak \citep[e.g.][]{mo96}.

Moreover, while these larger scales hold much potential for precise measurements of the shear signal, the modelling of this shear signal is more complicated (no longer an NFW profile) and cosmologically dependent. 
Furthermore, the neighboring halo is on average a low mass halo, and these are more strongly impacted by hydrodynamical effects. The mass distribution in the regime outside massive halos, but on much smaller scales than accurately predicted by linear theory, is 
an important source of systematic uncertainty for a variety of cluster cosmological experiments. 

To study the transition radius at which the cluster shear profile is no longer dominated by the 
central halo matter profile, we inspect the logarithmic slope of the density contrast $\Delta\Sigma$ stacked in mass bins for the different simulation snapshots in Fig.~\ref{fig:2halo}. We also plot a black vertical line at $R_\text{max}(z) =3.2/(1+z)\,h^{-1}\,\mathrm{Mpc}$. For all considered redshifts and masses, this line falls approximately at the minimum of the logarithmic slope. We identify this transition as the transition from the ``1-halo''-regime (where the central halo matter profile dominates the shear signal), to the ``2-halo'' term (which leads to a flattening of the shear profile). 

The redshift evolution of the transition radius is somewhat surprising. A distance that scales with redshift like $(1+z)^{-1}$ corresponds to a  fixed physical distance between two points that does not scale with the universal expansion.  This scaling might however well be coincidental.
Halos are gravitationally bound objects, detached from the background expansion. The radius enclosing a given over-density of matter grows with redshift as the density is diluted. Simultaneously, the amplitude of fluctuations in the Universe grows, increasing the 2-halo term. These two effects counteract each other. Further study of this interplay is surely warranted, but for our current purpose we can establish that limiting our fitting range to $R<3.2/(1+z)\,h^{-1}\,\mathrm{Mpc}$ excludes radial ranges not dominated by the central halo matter density profile. This avoids the complex cosmological dependencies and the modelling uncertainties of the 2-halo regime.

\subsection{Additional systematic uncertainties}\label{sec:further_sys}

\begin{table}
	\centering
	\caption{Estimates of the systematic uncertainty resulting from hydrodynamical modelling for different inner fitting radii $R_\text{min}$. These values are extracted from the comparison of a small number of independent simulations and should therefore be considered order of magnitude estimates. For the bias $\Delta b$ and the variance $\Delta s$, we use the maximum deviation found over different redshifts. The impact of hydrodynamical modelling uncertainties on the parameters of the weak lensing to halo mass mapping cannot be ignored.}\label{tab:hydro_sys}
	\begin{tabular}{llll}
	\hline
	$R_\text{min}$   & 0.2 $h^{-1}$Mpc  & 0.5 $h^{-1}$Mpc & 0.8 $h^{-1}$Mpc  \\
\hline
    $\Delta b$ & 0.06  & 0.02 & 0.03 \\
    $\Delta b_\text{M}$ & 0.016 & 0.018 & 0.017  \\
    $\Delta s$ & 0.15 & 0.25 & 0.2 \\
    $\Delta s_\text{M}$ & 0.26 & 0.59 & 0.60 \\
    \hline
	\end{tabular}
\end{table}

\begin{table}
	\centering
	\caption{Estimates of the systematic uncertainty resulting from the correlation between the cluster dynamical state and the strength of the mis-centering for different  inner fitting radii $R_\text{min}$. These values are extracted from the comparison of a two limiting case of very strong and very weak correlation, and should therefore be considered upper limits. For the bias $\Delta b$ and the variance $\Delta s$, we use the maximum deviation found over different redshifts.  Using larger inner fitting radii ensures that these uncertainties are smaller than other known uncertainties.} \label{tab:corr_miscentr_sys}
	\begin{tabular}{llll}
	\hline
	$R_\text{min}$   & 0.2 $h^{-1}$Mpc  & 0.5 $h^{-1}$Mpc & 0.8 $h^{-1}$Mpc  \\
\hline
    $\Delta b$ & 0.02  & 0.008 & 0.002 \\
    $\Delta b_\text{M}$ & 0.04 & 0.02 & 0.01  \\
    $\Delta s$ & 0.08 & 0.06 & 0.04 \\
    $\Delta s_\text{M}$ & 0.25 & 0.24 & 0.15 \\
    \hline
	\end{tabular}
\end{table}

While the numbers above provide the systematics budget for the most prominent systematic effects in a quantitative manner, some known systematics are left out because they are harder to quantify. These include the systematic uncertainties in the modelling of hydrodynamical effects in simulations and the impact of the -- possible -- correlation between the mis-centering and the dynamical state of the cluster.  
Therefore, we discuss our estimates for these systematic effects in the Appendices~\ref{sec:app_A} and \ref{sec:app_B} and provide a summary of our these estimates in Tables~\ref{tab:hydro_sys} and \ref{tab:corr_miscentr_sys}. 

At low reshift, the systematic uncertainty from  the treatment of hydrodynamical effects contributes to the error budget of the bias and its mass trend at a level that is comparable to other systematics. At high redshift, it is significantly smaller than the impact of photometric redshift biases (for the generic Stage III survey adopted here). Overall, we estimate that hydrodynamical effects contribute to a $\sim 2 \%$ systematic uncertainty floor on the weak-lensing mass bias. The uncertainty on the variance -- and its mass trend -- are clearly dominated by the modelling of hydrodynamical effects, independently on the choice of the inner fitting radius. This results in a systematic floor for the intrinsic scatter of weak-lensing mass estimates of about $\sim  10\%$. These estimates are quite crude as they rely on the comparison between two hydrodynamical simulations (Magneticum and Illustris TNG).  Further exploration of this uncertainty 
is clearly needed, especially in light of the 
goal of achieving 1\% WL mass accuracy in Stage IV surveys. These efforts would likely focus on 1) enlarging the number of hydrodynamical simulations to better probe the space of possible hydrodynamical effects, and 2) aiming to develop quantitative tools to assess the agreement between simulation predictions and observed data.

In contrast to the uncertainties from hydrodynamical modelling, the uncertainties stemming from the unknown degree of correlation between mis-centering and cluster dynamical state scales strongly with the inner fitting radius. This is to be expected, because mis-centering impacts the inner cluster regions much more. For the generic Stage III survey adopted here, these uncertainties can be reduced to a level that is smaller than the impact of other systematics by choosing a large inner fitting radius.

Among the systematic effects we did not discuss in this work, we find the impact of intra-cluster light on the photometric redshift estimation. \citet{gruen20} show that this effect is not relevant for current wide photometric surveys. We also did not consider the impact of blending on the shear measurement accuracy in crowded regions \citep{hernandezmartin20, sheldon20}. Both of these effects can be safely expected to be stronger in the cluster center, providing additional reasons to select a larger inner fitting radius.  

\subsection{Comparison of modeling approaches}\label{sec:hydro_approaches}

The systematic biases and associated uncertainties sourced by the impact of hydrodynamical effects on different quantities predicted by cosmological simulations have received increased attention over the last years.   Studies have been carried out on the impact of hydrodynamical effects on the matter power spectrum \citep[e.g.][]{white04, Jing06, guillet10, fedeli14, mead15, aschneider15, springel18, arico20}, the halo mass function \citep[e.g.][]{cui12, martizzi14, Velliscig14, bocquet16, castro21, Beltz-Mohrmann21} and the halo lensing signal \citep[e.g.][]{lee18, Debackere2021arXiv210107800D}. The latter two -- halo mass function and halo weak-lensing signal-- are of greatest interest to weak-lensing calibrated cluster number counts experiments, especially when the weak-lensing analysis is restricted to the inner, central halo dominated region (cf. Section~\ref{sec:outer_radius}).

Concerning the impact of hydrodynamical effects on the halo mass function, both \citet{castro21} and \citet{Beltz-Mohrmann21} constrain the relation between the halo masses in paired gravity-only and hydrodynamical simulations run with the same initial conditions. Simply adopting this approach for weak-lensing calibrated cluster cosmology would require one to calibrate the mapping between the halo mass in hydrodynamical simulation and the matter density profiles, which are themselves affected by the same hydrodynamical effects. This likely leads to correlations between the corrections applied to the halo mass function and the weak-lensing to halo mass mapping. This problem can be circumvented, and the overall set-up of the analysis simplified, if -- as we propose -- the mapping between the gravity-only halo mass and the hydrodynamics impacted weak-lensing shear signal is calibrated. While facilitating the cosmological inference from cluster number counts, this approach has the disadvantage of reducing the astrophysical information contained in inferred scaling relations, as they now relate to the abstract gravity-only mass, and not to the physically motivated halo mass, impacted by hydrodynamical effects.

Several recent works have suggested to absorb the impact of hydrodynamical effects on the weak-lensing shear signal by fitting for the concentration alongside the weak-lensing mass \citep{lee18, Debackere2021arXiv210107800D}. While this approach indeed produces less biased weak-lensing masses, it opens new challenges to the weak-lensing calibrated number counts experiments. Firstly, it requires one to marginalise over the concentration in the weak-lensing calibration likelihood. In our tests this significantly slows down the likelihood computation, making the analysis of O(100) clusters already computationally challenging. Furthermore, the signal-to-noise ratio on the concentration measurement for individual clusters is low, and the concentration is degenerate with the mass. In addition, we find that using flat priors on the concentration is not an uninformative choice, and leads to mass biases by artificially preferring a particular range of concentrations. On the other side, imposing informative priors requires accurate mass concentration relations. The accuracy of the mass--concentration simulations is likely going to be limited by the accuracy of hydrodynamical simulations, thus likely delivering no increase in the overall accuracy compared to the 
calibration of the weak-lensing to halo mass relation using a fixed concentration, as done here.

Considering this together with the increased computational cost of allowing concentration to float, it is clear that there are benefits of extracting weak-lensing masses with a model that has a fixed concentration or a fixed concentration--mass relation  \citep[see, e.g.][]{Mantz14, schrabback18a, dietrich19, stern19}.  Notably, \citet{Debackere2021arXiv210107800D} find that the bias between the weak-lensing mass and the gravity-only mass does not change significantly when fixing the concentration to the correct concentration--mass relation (Fig. 6, central panel), further justifying our approach. 

Generally, introducing additional observables (like fitting for the concentration, including the BCG orientation to estimate the cluster orientation \citep{herbonnet19}, or using BCG--X-ray offsets to estimate the amount of mis-centering) with the aim of increasing the precision (i.e. reducing the weak-lensing scatter $\sigma_\text{WL}$) 
has the risk of resulting 
in a loss of accuracy. The link between the additional observable and the underlying profile property would generally introduce additional systematic uncertainties. 
Hydrodynamical modelling uncertainties only slightly perturb the 2d projected matter density profile for massive halos, because the matter distribution is dominated by dark matter and it is projected along the line of sight. 
This reduces its sensitivity to baryon redistribution effects. Other cluster observables, like the BCG position, the X-ray center or cluster concentration, depend more strongly on the hydrodynamical modeling. 
Thus, they are inherently more challenging to calibrate with simulations. In the limit of an analysis that is systematics-dominated (large cluster samples, dense shear galaxy samples) it is crucial only to control the systematic uncertainties.  
On the other hand, in the limit of a small sample or a single cluster analysis or a large cluster sample with excellent control of systematic uncertainties, a higher fidelity model profile that employs additional information to reduce the scatter in the weak-lensing to halo mass relation could be advantageous.

\section{Conclusions}\label{sec:conc}

In this work we describe a method and provide a proof-of-concept application to quantitatively determine the
bias and accuracy of a weak-lensing mass calibration of a cluster sample. The concept of the \emph{weak-lensing} was previously used by \citet{becker11} to quantify the cluster-to-cluster variation w.r.t. to an NFW profile \citep[see also][]{Oguri2011MNRAS.414.1851O, Bahe12, lee18}.
Then, others followed \citep{applegateetal14, schrabback18a, dietrich19, stern19} who sought
to also account for the other systematic uncertainties in the weak-lensing measurement and folded these into the mapping between the weak-lensing and halo mass.
In this work, we systematize this concept, by explicitly marginalising over known weak-lensing systematics and using hydrodynamical simulations to simulate realistic shear profiles. By careful modelling choices, we ensure that the computational cost of the cosmological analysis using our framework is sufficiently small to allow us to analyse large cluster samples O(10,000) with individual cluster weak-lensing measurements.

Besides providing a clear procedure for deriving the uncertainties on the weak-lensing to halo mass, we make a 
conceptual improvement in the accounting of hydrodynamical effects in cluster number counts cosmology. Prior to this work, the impact of hydrodynamical effects on the abundance of halos and the impact of hydrodynamical effects of the matter profiles have been discussed separately.
Here we show how using suites of simulations run with and without hydrodynamical effects from identical initial conditions allows us to relate the gravity-only halo mass to the weak-lensing mass extracted in the presence of hydrodynamical effects. Using the calibration of this mapping, the cosmological dependence of the halo mass function extracted from large suites of gravity-only simulations can be exploited in a self-consistent manner that does not ignore hydrodynamical effects.

Our method starts from a library of 2d projected matter density profiles extracted from hydrodynamical simulations and the halo masses from the matched gravity-only runs. The projected matter density profiles are extracted for different mis-centered positions. We then specify a set of cluster properties and weak-lensing survey properties (such as an observable--mass relation, a multiplicative shear bias, an uncertainty on the lensing efficiency due to photometric redshift uncertainties, etc.) and transform the surface matter density profiles into shear profiles. A one-parameter model of the shear profile, inspired by a mis-centered NFW-profile, is then fit to the simulated shear profiles. The resulting best-fitting mass is called the weak-lensing mass. We fit the relation between weak-lensing and halo mass accounting for the mis-centering by appropriately weighing each mis-centered profile by the mis-centering distribution. This results in a set of weak-lensing to halo mass parameters.

We then repeat the extraction of the weak-lensing to halo mass parameters $\sim$1000 times with varying realisations of cluster and weak-lensing survey systematics. This  results in $\sim$1000 realisations of shear profile libraries sampling the entire space of weak-lensing systematics, and $\sim$1000 realisations of weak-lensing to halo mass parameters. The variance in this sample of weak-lensing to halo mass parameters encapsulates the systematic uncertainty on the weak-lensing mass measurement and can be used as a prior in cosmological analyses. Exploring the correlations between the weak-lensing to halo mass parameters and the cluster and weak-lensing systematics allows us to assess which systematic effects have the largest impact on the accuracy of the weak-lensing mass measurement.  For this exercise we adopt a Stage III like generic cluster survey modeled on the combination of SPT and DES.

We perform the calibration of the weak-lensing to halo mass relation for different inner fitting radii. We generally find that smaller inner fitting radii, including scales closer to the cluster center, lead to larger systematic uncertainties in the weak-lensing mass measurement. This loss in accuracy is fractionally larger than the gain in precision due to the inclusion of more, stronger sheared background galaxies.  Selection of an optimal inner radius depends on the details of a cluster weak-lensing experiment, but in the limit of analyses of large cluster samples that are systematics-dominated, there are clear advantages in avoiding the cluster core region.

Inspecting the logarithmic slope of the shear profile, we determine the outer fitting radius $R_\text{max}=3.2\,(1+z)^{-1}\,h^{-1}\,\mathrm{Mpc}$ which excludes the signal from correlated neighboring structures. That signal -- the so-called 2-halo term -- has complex dependencies on cosmological parameters and hydrodynamical effects, and 
for that reason we have excluded it from our analysis.

Finally, we also provide estimates for the uncertainty stemming from the hydrodynamical modelling and from the possible correlations between mis-centering and cluster dynamical state. The former leads to a $\sim 2\%$ systematics floor, comparable to other systematics for Stage III surveys, but clearly above the $1\%$ goal for Stage IV surveys.  
Further reducing this uncertainty would require that a wide range of hydrodynamical models fully sampling the plausible range of sub-grid physics be carried out and then compared to observational data to assess the likelihood of each hydrodynamic model \citep[e.g.][]{chisari19, arico20}.  That ensemble of simulations and the associated likelihoods could then be used to reduce the hydrodynamics related systematics floor.

In summary, this work discusses a new framework for deriving the systematic uncertainty on the weak-lensing mass measurements for cluster surveys. The proposed method -- shown in the context of a generic Stage III cluster and weak-lensing survey -- accounts for the impact of hydrodynamical effects, different data systematics and marginalizes over mis-centering effects, while ensuring an efficient mass extraction that is well-suited for large cluster surveys. Our method is being applied to the weak-lensing mass calibration of South Pole Telescope selected clusters with Dark Energy Survey weak-lensing data \citep{bocquetip}, and the weak-lensing mass calibration of  clusters selected in the eROSITA full survey depth equatorial field with Hyper-Supreme Cam weak-lensing data \citep{chiuip}.
Our framework can naturally be extended to incorporate the properties of upcoming Stage-IV lensing surveys such as those by Vera C. Rubin Observatory \citep{Ivezic08} and Euclid \citep{laureijs11}.

\section*{Acknowledgements}

It is our pleasure to thank Antonio Ragagnin for his support with \textsc{g3read}\footnote{\url{https://github.com/aragagnin/g3read}}, a python tool to read Gadget files such as those produced by Magneticum.
We would also like to thank Martin Sommer and Tim Schrabback for providing their results on the weak-lensing bias and scatter for centered shear profiles for a successful cross check, and Anja von der Linden for comments on extraction from simulations. We would like to thank the members of the Euclid, SPT, and DES
Cluster Working Group, and of DESC-Clusters for the useful discussions on the draft of this paper.

We acknowledge the support by the ORIGINS Cluster funded by the Deutsche Forschungsgemeinschaft (DFG) under Germany's Excellence Strategy -- EXC-2094 -- 390783311, together with the support of the Max Planck Society Faculty Fellow program and the Ludwig-Maximilians-Universitaet.

\section*{Data Availability}
The data underlying this article will be shared upon 
request to the corresponding author.


\bibliographystyle{mnras}
\bibliography{example} 


\appendix

\begin{figure*}
	\includegraphics[width=\textwidth]{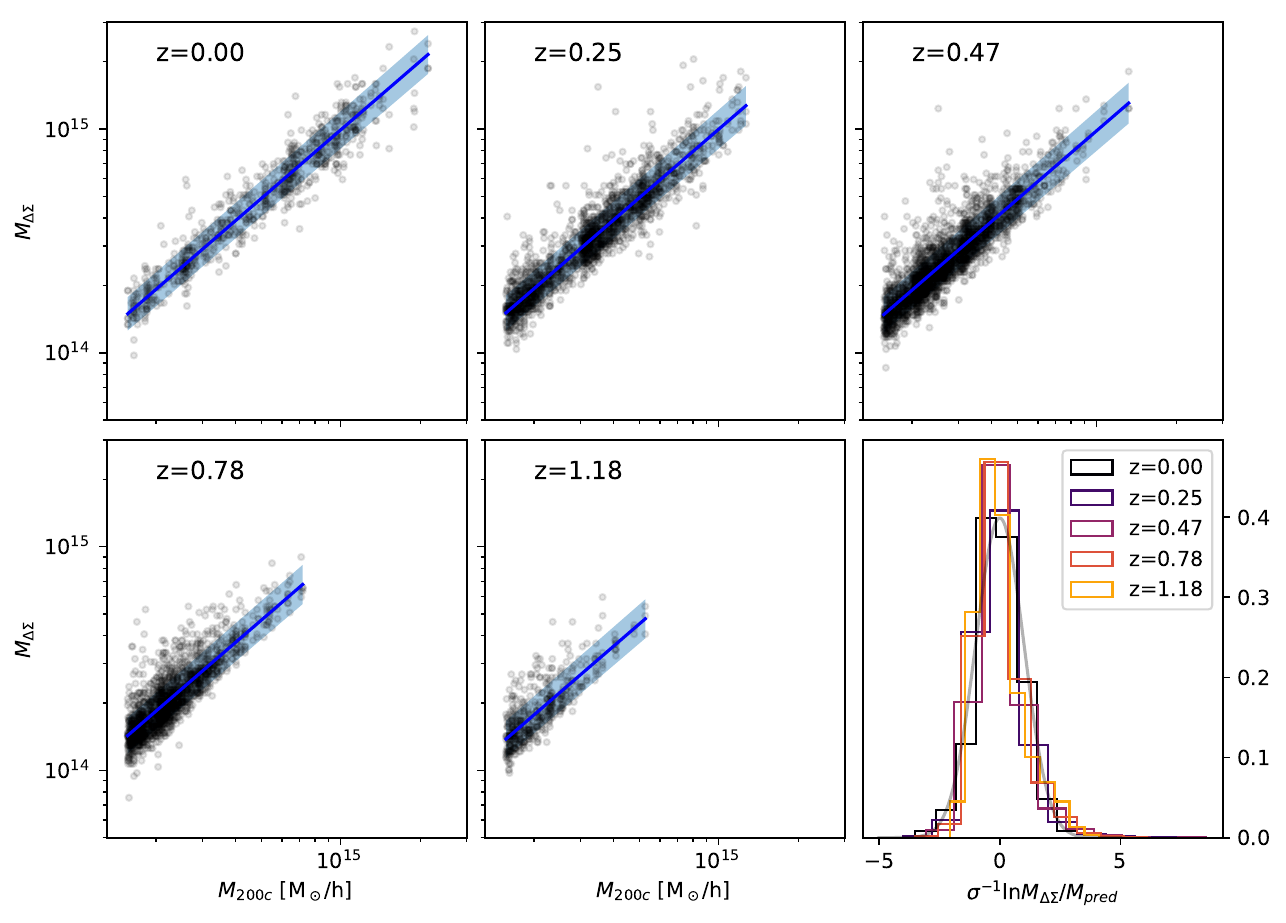}
	\vskip-0.10in
    \caption{\textit{First 5 panels:} Mass $M_{\Delta\Sigma}$ (extracted from fitting an NFW profile to the perfectly centered density contrast profiles) plotted against the halo mass $M_\text{200c}$ for halos extracted from different snapshots (different panel) in the Magneticum hydrodynamical simulation, together with a log-normal fit to the distribution as a blue line (intrinsic scatter in shaded blue).  \textit{Lower left panel:} Residuals of the fits in the different snapshots, normalised by the intrinsic scatter. In gray a standard normal distribution. Some positive skewness in the mass distribution is noticeable. It might be related to line of sight project effects. Comparing the $M_{\Delta\Sigma}$-- halo mass relation from different simulation provides an estimate on the residual uncertainty stemming from the treatment of hydrodynamical effects.}\label{fig:mdeltasigma_vs_mhalo}
\end{figure*}

\begin{figure}
	\includegraphics[width=0.92\columnwidth]{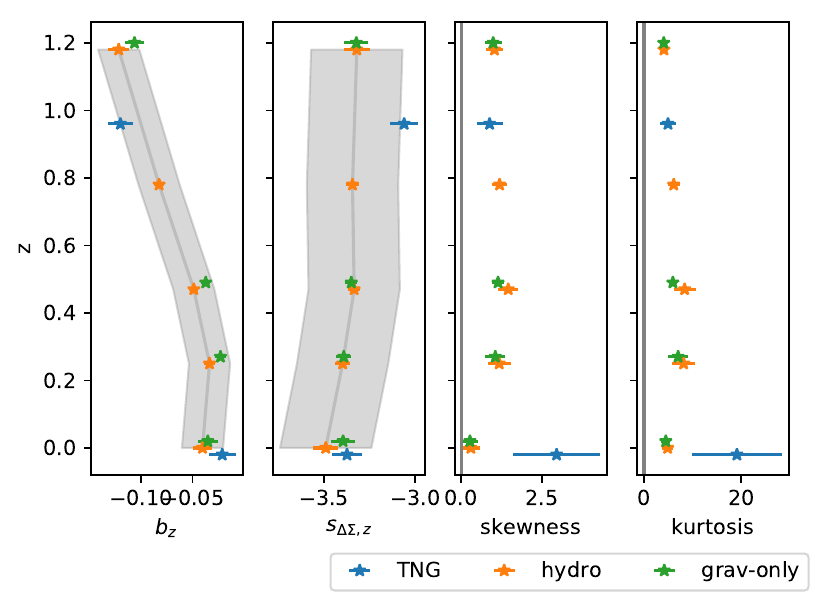}
	\vskip-0.10in
    \caption{Bias between the density contrast mass and the halo mass, natural logarithm of  the variance of  the the density contrast mass, $s_{\Delta\Sigma} = \ln \text{Var}[\ln M_{\Delta\Sigma} | M_\text{200c}]$, and skewness and kurtosis of the distribution of density contrast masses around the halo masses for different simulations at different redshift. The gray band indicate the maximal difference between any simulation and Magneticum hydro. We use this quantity to estimate the systematic error introduces by the modelling of the hydrodynamical effects (gray band). In the two rightmost panels, the gray lines indicate the reference value for Gaussian distributions (skewness=0, kurtosis=3). We find a positive skewness and large kurtosis in all simulations.}\label{fig:results_MDeltaSigma}
\end{figure}

\section{Systematic budget for hydrodynamical modelling}\label{sec:app_A}

Each simulation we use has a specific set of assumptions for the hydrodynamical modelling. As such, each simulation is a point estimate of the hydrodynamical modelling uncertainties. Despite ongoing efforts, it is to date very costly to  sample the entire range of possible hydrodynamical modelling systematics (several thousand simulations would have to be run).  In this work, we bracket the remaining uncertainty by determining the difference in weak-lensing to halo mass scaling relation parameters between three different simulations: Magneticum with hydrodynamical effects, Magneticum gravity-only, and Illustris TNG with hydrodynamical effects.


For simplicity, we specifically look at centered density contrast profiles $\Delta\Sigma(R)$ and extract the density contrast mass $M_{\Delta\Sigma}$ by fitting an NFW profile with constant concentration ($c=3.5$) to the density contrast profile of each halo. We use a minimal fitting radius of 0.5 $h^{-1}$Mpc, and an outer fitting radius following the prescription discussed above. The scatter plot between the halo mass and the density contrast mass in the Magneticum hydrodynamical run is shown in Fig.~\ref{fig:mdeltasigma_vs_mhalo}.  We then fit a log-normal distribution with mass trend in the mean and the variance (as specified in Eq.~\ref{eq:meanMWLgivenM}, Eq.~\ref{eq:PlnMWLgM},  and Eq.~\ref{eq:varMWLgivenM}). This results in bias and intrinsic scatter parameters shown in the two leftmost panels of  Fig.~\ref{fig:results_MDeltaSigma}. On the Magneticum hydro simulations we find a mass trend of the bias of $b_\text{M}=1.019\pm0.006$, and a mass trend of the variance $s_\text{M}=0.11\pm0.04$. 

The same analysis is repeated also for the Magneticum gravity-only simulations, and for the Illustris-TNG simulations. The resulting biases and intrinsic scatters are shown in the two leftmost panels of  Fig.~\ref{fig:results_MDeltaSigma}. For Magneticum gravity-only  (Illustris-TNG) simulation we find a mass trend of the bias of $b_\text{M}=1.001\pm0.006$ ($1.037\pm0.012$), and a mass trend of the variance $s_\text{M}=0.28\pm0.04$  ($-0.48\pm0.08$). We then estimate the systematic uncertainty induced by different -- or no -- hydrodynamical treatment by taking the difference between the parameters from Illustris-TNG and the parameters from Magneticum hydro. Note that this difference is always larger than the difference between Magneticum hydro and Magneticum gravity-only. The resulting estimates are shown in Table~\ref{tab:hydro_sys}  for different inner fitting radii (the same analysis as above is repeated also for $R_\text{min}=0.2,\,0.8\,h^{-1}\,\mathrm{Mpc}$). Note that these numbers are estimated from a very limited number of simulations. As such, they are to be considered order  of magnitude estimates. The three simulations we compared also have quite different universal baryon fractions: 16.8$\%$ for Magneticum--hydro, 15.7$\%$ for Illustris-TNG--hydro. This impacts the strength of the hydrodynamical feedback. Our estimated uncertainties include this effect, making them conservative estimates. Proper marginalisation over a sufficiently wide range of hydrodynamical modelling and cosmological parameters currently eludes our possibilities, due to the large computational cost.

At low redshift, the hydrodynamical  modelling  introduces as error on the biases that is comparable in magnitude to the uncertainty derived from the better quantified systematics (compare Table~\ref{tab:results} and Table~\ref{tab:hydro_sys}). At higher redshift, the large uncertainties introduced by the photometric redshift uncertainty surpass the error resulting from the  hydrodynamical modelling. In the case of the intrinsic variance, the error budget from hydrodynamical modelling is rather constant with inner fitting radius, while the effect from  the other systematics is reduced strongly when using a larger inner fitting radius. At $R_\text{min}=0.8\,h^{-1}\,\mathrm{Mpc}$, the effects of hydrodynamical modelling are the dominant source of systematic uncertainty. In the case of  the mass trend of the bias, the uncertainty from hydrodynamical modelling is of the same order of magnitude as the effect of other systematics. In contrast, the uncertainty on the mass trend of the variance is dominated by the impact of hydrodynamical modelling. In summary, hydrodynamical modelling contributes strongly to the uncertainty on the variance and its mass trend, while it plays an important role in the systematics budget of the bias and its mass trend. We add these contributions in quadrature to the errors resulting from the marginalisation over the other systematics.

\section{Extra systematics related to mis-centering}\label{sec:app_B}

Our treatment of mis-centering rests on some rather strong assumptions. We make assumptions on the direction of the mis-centering and on the correlation between its strength and the cluster dynamical state. We are going to estimate the impact of the latter on the uncertainty of the weak lensing bias and scatter.

\subsection{Direction of the mis-centering}
We assigned the offset isotropically in the plane of the sky. Depending on the sort of mis-centering one plans to emulate, this assumption might be wrong. If the mis-centering is due to a physical reason (physical offset between brightest central galaxy (BCG) and halo center, or physical offset between peak in the X-ray surface brightness and halo center, etc.) it stands to reason that such an offset would correlate with the triaxial structure of the halo: it would be more pronounced along the major axis, and less pronounced along the minor axis. How strong this effect actually is, is yet to be determined empirically -- and in our opinion very hard to measure. 

Similarly, the mis-centering can be due to mis-association or other forms of ambiguity. The automated codes could have selected a galaxy far from the cluster center as the BCG, or the system  could be undergoing a merger, offsetting its gas component from the stellar and dark matter components. In such cases, the matter profile around the observed center might be very different from that of a relaxed cluster, artificially mis-centered by an equal amount. Qualitatively, such systems only make a few percent of typical cluster samples.

Finally, a 2d isotropic mis-centering is actually a correct description of the mis-centering due to the measurement uncertainty induced by survey instruments with a comparatively bad angular resolution (typically X-rays or SZ surveys). In this case, noise fluctuation scatter the measured center around the true center in an isotropic manner on the plane of the sky.

In summary, the  correct directional assignment of the mis-centering is highly dependent on which centers one uses for the weak-lensing analysis. There remain also many uncertainties and unknowns concerning the quantitative description of the direction of the mis-centering. In our opinion, this provides an additional reason to choose large inner fitting radii.  

\subsection{Correlation between strength of mis-centering and cluster dynamical state}

The second assumption we made, is that the strength of the mis-centering (as  parameterized by Eq.~\ref{eq:P_Rmis}) does not correlate with any cluster property. A quantitative description of much cluster centers correlate with cluster properties is, however, both highly dependent on the type of centres one chooses, and also very uncertain, as -- to the authors' knowledge -- no empirical study in the context of large surveys has found a conclusive measurement of such correlation. 

In order to bracket the possible impact of such a correlation, we construct a case in which the strength of mis-centering and the cluster dynamical state are highly correlated. For each of the 3 projections of a halo, we measure the concentration by fitting a two parameter NFW-model to the centered density contrast. We then rank all halo-projections based of their 2d concentration. The $\rho$ steepest halo-projections then get assigned the narrower component of the mis-centering distribution, while the $1-\rho$ flattest halo-projections get assigned the larger component of the mis-centering distribution. This way, the question whether the halo is strongly or weakly mis-centered along that projection is made to correlate perfectly with concentration of the density contrast profile in that projection. 

Operationally, this requires us to alter the mis-centering distribution used in the fitting of the weak-lensing -- halo mass parameters (Eq.~\ref{eq:likeWLmass-mass}). In this case, it depends on the 2d concentration. The extraction of the weak-lensing mass is kept agnostic to the cluster 2d concentration, as in most large survey applications, the signal-to-noise of cluster-by-cluster weak-lensing measurements is too low to measure the concentration at all (considering that in many cases, even the mass measurement is only about 1-2 $\sigma$ with fixed concentration). 

This analysis then results in a new set of bias and variance parameters for the weak-lensing -- halo mass relation. In table~\ref{tab:corr_miscentr_sys} we report the absolute difference between the baseline results and the results where the strength of the mis-centering is perfectly correlated with the 2d concentration of the halo-projection. We perform the analysis for different inner fitting radii. We find that using large inner fitting radii reduces the difference to a level clearly smaller than the systematic uncertainty from other sources.


\bsp	
\label{lastpage}
\end{document}